\input epsf
\documentstyle[eqsecnum,floats,aps]{revtex}
\begin{document}
\title{Head-on collision of two black holes:\\
comparison of different approaches}

\author{Peter Anninos${}^{1,2}$, Richard H. Price${}^{3}$,
Jorge Pullin${}^{4}$, Edward
Seidel${}^{1,2}$, and Wai-Mo Suen${}^{5}$}

\address{
$^{1}$ National Center for Supercomputing Applications, \\
Beckman Institute, 405 N. Mathews Ave., Urbana, IL, 61801 \\
$^{2}$ Department of Physics,
University of Illinois, Urbana, IL 61801 \\
$^{3}$ Department of Physics, University of Utah, Salt Lake City, UT
84112-1195 \\
$^{4}$ Center for Gravitational Physics and Geometry, Department of Physics,
104 Davey Lab,\\
The Pennsylvania State University, University Park, PA 16802,\\
$^{5}$ McDonnell Center for the Space Sciences, Department of Physics, \\
Washington University, St. Louis, Missouri, 63130
}
\date{\today}
\maketitle
\begin{abstract}
A benchmark problem for numerical relativity has been the head-on
collision of two black holes starting from the ``Misner initial
data,'' a closed form momentarily stationary solution to the
constraint equations with an adjustable closeness parameter
$\mu_0$. We show here how an eclectic mixture of approximation methods
can provide both an efficient means of determining the time
development of the initial data and a good understanding of the physics
of the problem.  When the Misner data is chosen to correspond to holes
initially very close together, a common horizon surrounds both holes
and the geometry exterior to the horizon can be treated as a
non-spherical perturbation of a single Schwarzschild hole. When the
holes are initially well separated the problem can be treated with a
different approximation scheme, ``the particle-membrane method.''  For
all initial separations, numerical relativity is in principle
applicable, but is costly and of uncertain accuracy.  We present here
a comparison of the different approaches.  We compare waveforms, for
$\ell=2$ and $\ell=4$ radiation, for different values of $\mu_0$, from
the three different approaches to the problem.
\end{abstract}
\vspace{-12cm}
\begin{flushright}
\baselineskip=15pt
CGPG-95/3-6  \\
gr-qc/9505042\\
\end{flushright}
\vspace{13cm}
\vskip 1.5cm
\twocolumn
\section{Introduction}
\label{sec:intro}

The advent of LIGO and VIRGO \cite{Abramovici92}, and other
gravitational radiation detectors in the coming years, adds
motivation to achieve a better understanding of the physics
of black hole collisions, physics that is already interesting
for many other reasons. Not the least of these is that black hole
collisions were the earliest testbed for numerical relativity,
the solution of the differential equations of Einstein's theory
with numerical codes. Smarr and Eppley\cite{Smarr75,Eppley75,Smarr79},
more than 15 years ago, computed the radiation waveforms for
the axisymmetric problem of two holes, starting from rest
and falling into each other in a head-on collision.
Partly to gauge the progress of the past 15 years, both
in computing machinery and in its application to the problem,
the problem has recently been reconsidered and
recomputed\cite{Anninos93b,Anninos94b}.  One of the reasons for the
importance of this problem is that it is a starting point for a major
initiative in state-of-the-art computing, the Grand Challenge
project\cite{GC94} to compute the spiraling coalescence of binary
black hole systems in the absence of symmetries.  This project will
rely on both large scale numerical computations to solve the fully
nonlinear Einstein equations and on a new set of semi-analytic techniques,
such as those presented here, to gain a physical understanding of
the numerical results and to provide testbed calculations to verify
the new generation of numerical codes under development\cite{Anninos94c}.

The technical difficulties of solving the dynamics of black hole
interactions are daunting, as will be clear in the limits of progress
reported below. The difficulties make it important, or even necessary,
that the numerical work be supported by other approaches.
In particular, the numerical challenge would be greatly aided by
analytic or semi-analytic methods of computing black hole solutions
in cases where such methods can be devised, so that those developing
numerical codes would be guided.  Most important, distinctly different
methods of finding solutions can provide a measure of the errors,
especially the systematic errors, in the numerical schemes.

There is a very separate motivation for developing alternative methods
of finding solutions. While numerical treatment gives us the data
representing the physical process, analytical methods provide
a structure for understanding these data, for perceiving what
is interesting, what is expected and is new, and what generalizations
are plausible, and what questions should be asked next.

We report here on work done with just such motivations. The problem is
the head-on collision of two holes, the classic problem of Smarr and
Eppley. The initial data is the analytical solution of the Einstein
initial value equations given by Misner\cite{Misner60}. This solution
represents two symmetric, momentarily stationary, ``throats'' with a
proper separation $L$, in a spacetime with ADM mass $M$ (thus
representing two momentarily stationary holes each of mass roughly
$M/2$)\cite{beware}. In this solution the dimensionless measure of
initial separation $L/M$ of the throats is given by a parameter
$\mu_0$.

Initial data with large values of $\mu_0$ represent infall scenarios
starting from large separations. Small $\mu_0$
corresponds to very different collapses starting with close throats.
Three distinctly different methods are used to study the problem:
(i)~Over a broad range of separations, the more-or-less established
approach to numerical relativity, the numerical solution of the
Einstein equations in 3+1 form, is used to find the spacetime evolved
{}from the initial data. (ii)~For large separations (``far limit'') we use
the ``particle-membrane'' method, which starts with a point
particle infall treated using perturbation theory, and
then introduces factors describing the internal dynamics of the black holes,
using the black hole membrane paradigm.  (iii)~In the
``close limit,'' when initial separations are small, we exploit the fact that
the initial geometry is nearly spherical outside the initial horizon;
the evolved spacetime outside the horizon is therefore nearly
spherical and its development can be approximated with the theory of
non-spherical perturbations of a single black hole.

The range of validity of the three methods is shown below and their
strengths and weaknesses discussed. The conclusion that emerges is
that the eclectic approach to this problem, using a mixture of
distinctly different computational methods, gives a very robust
set of answers, with good limits on errors, and much improved
understanding of the meaning of the answers. We argue that similar
approaches should be developed, wherever possible, as the forefront
of the Grand Challenge initiative moves forward.

The remainder of the paper is organized as follows: In Sec.~II an
outline is given of how the problem is translated into a problem
suitable for numerical computation, and how information is extracted
about the outgoing radiation.  The ``particle-membrane'' method
suitable for large initial separations is discussed in Sec.~III and
numerical results from this method are presented. In Sec.~IVA an
introduction is given to the ``close limit,'' the techniques of
perturbation theory applied to collisions with small initial
separations. In Sec.~IVB numerical results of the close limit are
shown and are compared with the methods of Sec.~II and ~III.
Conclusions are briefly presented in Sec.~V.

\section{Numerical Relativity for the Head-on Black Hole Collision}
\label{sec:numrel}

We use the standard $3\!+\!1$ (ADM) formalism as the framework
for building a numerical code to solve the fully nonlinear
Einstein equations, and evolve the Misner initial data sets
without making any approximations.  The \v{C}ade\v{z} coordinate
system~\cite{Cadez71} is used, for the most part, as it provides
natural spherical boundaries at the black hole  throats and in
the asymptotic wave zone where radiation is extracted. It also
utilizes a logarithmic ``radial'' coordinate to extend grid
coverage beyond where radiation can propagate within a typical
run time.  However, this coordinate system has a singular saddle
point midway between the two holes, making numerical evolution
quite difficult.  For this reason an overlapping cylindrical
coordinate system is used as a coordinate patch to cover this
region, independently of the \v{C}ade\v{z} grid except at the interface
boundaries.  The nonsingular cylindrical metric and extrinsic
curvature components are then used to correct the corresponding
singular \v{C}ade\v{z} components in the overlapping patched regions
throughout the evolution.  We mention this detail because although
this procedure was very effective in suppressing numerical
instabilities that can develop at the singular point, it also
has the effect of introducing low amplitude signals in the
evolution that have a bearing on the interpretation of the
gravitational radiation waveforms presented below.
This work has been discussed extensively in
Refs.~\cite{Anninos94b,Anninos93a,Anninos94a},
where complete details of our numerical calculations
and results can be found.  Here we focus the discussion
on the extraction of waveforms from the numerically generated
spacetime metric.  In later sections of this paper, we compare
the waveforms extracted to those obtained using the semi-analytic
approaches.

Our waveform calculations are based on the gauge invariant extraction
technique developed by Abrahams and Evans~\cite{Abrahams90}
and applied in Ref.~\cite{Abrahams92a} to black hole spacetimes.
This work, in turn, is derived from the gauge-invariant formalism
developed by Moncrief~\cite{Moncrief74}.  The basic idea is
to split the numerical spacetime metric into a spherically
symmetric background and a small non-spherical perturbation.
First, we expand the metric perturbation in $m=0$ spherical
harmonics $Y_{\ell 0}(\theta)$ and their tensor generalizations.
Then, as described in Appendix~A of Ref.~\cite{Abrahams92a}, Moncrief's
perturbation functions $H_2$, $G$, and so forth, are
projected out of the numerically computed metric components
using the orthogonality of the $Y_{\ell 0}(\theta)$'s by
performing angular integrals.  For example, for each $\ell$--mode,
if the background is written in Schwarzschild coordinates,
the perturbation function $H_2$ can be computed by performing
an integral of the numerically evolved radial metric function
$g_{rr}$ over a 2--sphere
\begin{equation}
  H_2^{(\ell)}
= 2\pi \left( 1-\frac{2M}{r}\right)
  \int_0^{\pi} g_{rr} Y_{\ell 0}(\theta) \sin\theta d\theta \quad.
\end{equation}
The other perturbation functions are computed in a similar way.
These perturbation functions will be discussed in more detail
below in section \ref{sec:pert} where they are computed in
a different way using analytic approximation techniques.

Once these perturbation functions are known numerically,
they are used to construct the gauge invariant Zerilli
function $\psi$.  Following Moncrief~\cite{Moncrief74}
we define the following quantities
\begin{eqnarray}
  k_1 \equiv && K + Sr G,_r - 2\frac{S}{r} h_1 \quad, \label{k1eq}\\
  k_2  \equiv && \frac{H_2}{2S} - \frac{1}{2S^{\frac{1}{2}}}
       \frac{\partial}{\partial r}
       \left(rS^{-\frac{1}{2}}K\right)\quad,\label{k2eq}
\end{eqnarray}
where $S=1-2M/r$.  Then the quantity defined by
\begin{equation}\label{psi eq}
      \psi_{\rm num}
\equiv \sqrt{\frac{2(\ell -1)(\ell +2)}{\ell (\ell +1)}} \,
       \frac{\left[4rS^2k_2+ \ell (\ell + 1) rk_1\right]}{\Lambda}\quad,
\end{equation}
where
\begin{equation}
  \Lambda \equiv \left[\ell (\ell +1) -2+\frac{6M}{r}\right]\quad,
\end {equation}
is gauge invariant and satisfies the Zerilli equation describing
the propagation of gravitational waves on a black hole background.
With our choice of normalization given by Eq.~(\ref{psi eq}),
the asymptotic energy flux carried by the gravitational wave
is given by
\begin{equation}
 \frac{dE}{dt}
= \frac{1}{32 \pi} \left({\frac{\partial \psi}{\partial t}} \right)^2
\end{equation}
for each $\ell$ mode.

It is important to note that this procedure for extracting the Zerilli
function $\psi$ differs slightly from the procedure used in
Sec.~\ref{sec:pert}.  Here we imagine the spacetime at a large
distance to be well represented by spherical plus non-spherical pieces,
corresponding to various $\ell$--poles.  Although we assume such
nonspherical pieces to be small enough that nonlinear terms in the
non-spherical pieces can be neglected with impunity, the procedure
described above simply lumps ``everything not spherical'' into the
perturbation terms, whether it is small or not.  In this sense our
numerically extracted waveforms could be considered ``nonlinear.''
In the semi-analytic approximation methods discussed
below higher-order terms are explicitly dropped everywhere in
the spacetime, so that only the ``true'' first-order part is included.
Of course we expect the methods to agree in the regime where
the approximations we use are valid, such expectations are
borne out as we show in section~\ref{sec:results}.

We have extracted both the $\ell = 2$ and $\ell = 4$ waveforms, using
the method described above, at radii of 15, 20, 25, 30, and $35\,M$,
where $M$ is the ADM mass of the spacetime.  By comparing results
at each of these radii we are able to check the propagation of waves
and the consistency of our energy calculations.  Possible sources
of numerical errors in the gravitational wave signals include:
(i)~truncation error, (ii)~artificial diffusion,
(iii)~grid spacing that increases exponentially with ``radius'',
(iv)~severe pathological behavior arising from the singularity
avoiding (maximal) time slicing, and (v)~coordinate patch
implementation. These effects have been discussed in detail in
Ref.~\cite{Anninos94a} where we have performed a number of convergence
studies  and demonstrated the robustness of the $\ell=2$ waveform
extraction to a few percent under significant changes in computational
parameters such as grid resolution, patch width, numerical
diffusion, etc.  The most prominent numerically induced
feature in the $\ell=2$ waveform is a slight broadening of
the wavelength at late times.  The $\ell=4$ waveform is less
certain than $\ell=2$ (especially at low values of $\mu_0$)
due to its much smaller amplitude which makes it difficult
to extract from the background noise level.  Moreover, the more
complicated angular distribution of the $\ell=4$ component requires
more angular zones to be fully resolved than the $\ell=2$ component.
Our convergence studies indicate that although we are unable
to determine with great certainty the absolute amplitude of
the $\ell=4$ signal, the damping rate, frequency and phase
of these waveforms are significantly more reliable.
We shall see the confirmation of these points, together with new
insights into the numerical data, from the comparison with
the results from the semi-analytic approaches.

\section{The Particle-Membrane Approach}

In this section we consider the case in which the two
black holes are initially well separated --- the ``far limit.''
In this case the evolution of the system can be separated into two
phases. The first is the infall phase, during which the
radiation emitted depends almost uniquely on the motion of
the two bodies, and almost not at all on their nature (that is,
whether or not they are black holes). The second is the
interaction phase,
during which the radiation emitted can depend
significantly on the nature of the two bodies; it is during this
phase that most of the gravitational-wave energy is produced.

We describe a semi-analytic approach, first presented in
Ref.~\cite{Anninos94b}, for calculating the gravitational
radiation produced in the head-on collisions described
by the Misner data. The approach
consists of two parts. In the first (particle) part,
we calculate the radiation by assuming that only one body is
a black hole, and that the other is a much less massive
(structureless) point particle; then we extrapolate the
results to allow for a mass ratio of unity. The first part of
the calculation incorporates completely the infall phase of
the collision, but only partially the interaction phase. In the second
(membrane) part of the calculation, we promote the
point particle to a black hole, and compute corrections
due to the internal dynamics of that black hole. These
corrections are important only during the interaction phase
of the evolution. The first part of the calculation is carried
out using black-hole
perturbation theory; the second is carried out using the
membrane paradigm for black-hole dynamics.

The particle-membrane approach was used in
Ref.~\cite{Anninos94b} to estimate the energy
carried off by gravitational radiation.
The result was found to be in
excellent agreement with the numerical calculation;
see Fig. \ref{fig3.1} in Sec.~III A.  Here we push the comparison
one step further: We use the particle-membrane approach
to calculate the waveforms, which we then compare to
those obtained numerically.

In Sec.~III A (infall from infinity) we construct the
waveform by considering a particle falling in toward the
black hole from infinity.
We shall see that
although at late times the agreement
is very good, the waveform differs from the numerical one
by some low frequency components at earlier times.
This is to be expected, because the physical
situations considered are very different from each other at early
times:
infall from infinity versus infall from a finite distance.
These differences in waveforms are nevertheless compatible
with a good agreement in the total energy outputs, as
the early part of the infall does not generate much radiation.

In Sec.~III B (time-symmetric trajectory) we compute improved
waveforms based
on particle motions bearing a closer resemblance to the numerical
set-up. We consider a particle emerging out of the white-hole (past)
horizon, coming to a stop at some finite distance away from the black
hole, and then falling back.  The waveforms
obtained are compared to the numerically generated ones in detail.

\subsection{Infall from Infinity}

We begin by outlining the particle-membrane calculation presented
in Ref.~\cite{Anninos94b}. The emphasis will be put on those
steps which are relevant for the computation of the waveforms.

The starting point is the twenty-three-year old computation
\cite{DRPP} of the energy radiated by a particle falling
into a black hole from infinity. The particle has mass $m_1$,
while the mass of the (non-rotating) black hole is denoted $m_2$.
Under the assumption $m_1 \ll m_2$, the total energy radiated can be
calculated using black-hole perturbation theory \cite{DRPP}.
The result is
\begin{equation}
  E = \kappa {m_1}^2/m_2, \qquad \kappa = 0.0104.
\label{3.1}
\end{equation}
To compare this with the numerical results, we must first
(i) extrapolate to a mass ratio of unity, (ii) take into
account the fact that in the numerical simulations, the
infall proceeds not from infinity but from a finite distance,
and (iii) correct for the internal dynamics of the infalling
black hole.

To extrapolate to the case $m_1=m_2$, we begin
by first understanding the origin of the factor ${m_1}^2/m_2$ in
Eq.~(\ref{3.1}). This, we do by invoking the quadrupole
formula for gravitational radiation.
If $I$ denotes a typical component of the system's quadrupole
moment, then
\begin{equation}
I \sim m_1 r^2,
\end{equation}
where $r$ is the distance to the
origin (where $m_2$ is assumed to sit).
The gravitational-wave luminosity $L=dE/dt$ is then given by
\begin{equation}
L \sim \left(\frac{d^3 I}{dt^3} \right)^2 \sim
\left( m_1 \frac{dr}{dt} \frac{d^2r}{dt^2} \right)^2 \sim
\frac{{m_1}^2 {m_2}^3}{r^5}.
\label{ll}
\end{equation}
To obtain the total energy radiated, we integrate this over time.
The dominant contribution comes from the strong-field region, so
\begin{equation}
E \sim L_{\mbox{strong field}}\, \Delta t_{\mbox{strong field}},
\label{ee}
\end{equation}
where $\Delta t_{\mbox{strong field}}$ is the time required by the
particle to cross the strong-field region, which has an extension
$r_{\mbox{strong field}} \sim m_2$,
so that $\Delta t_{\mbox{strong field}} \sim m_2$.
Putting these into Eq.~(\ref{ll}) and Eq.~(\ref{ee})
yields $E \sim {m_1}^2/m_2$, as in Eq.~(\ref{3.1}). In particular,
we
point out that $E$ is inversely proportional to $m_2$, the
length scale associated with the strong-field region.

We now repeat this argument having in mind an infalling
particle which has a mass $m_1$ not negligible compared
with the mass $m_2$ of the black hole.
The quadrupole moment must now be replaced
by $I \sim \mu r^2$, where $\mu = m_1 m_2/(m_1+m_2)$ is the
reduced mass and $r$ the relative separation.
On the other hand, quite
independent of the value of $m_1$, the extension
of the strong-field region as seen by the infalling particle
is still $r_{\mbox{strong field}} \sim
\Delta t_{\mbox{strong field}} \sim m_2$.  Hence Eq.~(\ref{ee})
now leads to
$E \sim \mu^2/m_2$. Finally, taking the limiting case
$m_1=m_2$, and inserting the numerical factor
$\kappa$ from  Eq.~(\ref{3.1}), we arrive at
\begin{equation}
E = 2 \kappa \mu^2 / M = \kappa M/8,
\label{3.2}
\end{equation}
where $M = m_1 +m_2 $ is the total mass of the system.  Notice that,
say, if one replaces $m_1$ by $\mu$, but $m_2$ by $M$ in (3.1), the
result will be different by a factor of 2.  We note that while
there is no definite way to carry out the extrapolation,
 to the extent that a factor of 2 is meaningful in such
arguments, our way of doing it is physically motivated.  We discuss
this point in detail here as in a different consideration below, $m_2$
will be extrapolated to $M$ instead of $M/2$.

Next, we modify Eq.~(\ref{3.2}) to reflect the fact that in
the numerical simulation, the infall proceeds from a finite
initial separation (denoted $r_0$) instead of infinity.
This modification was described in detail
in Ref.\cite{Anninos94b}, and we will not repeat this
discussion here. Suffice it to say that an infall from
a finite distance produces less radiation than an infall
{}from infinity. This is because (i) there
is less time for the system to radiate, and (ii) the
infall velocity is smaller at a given separation. The
reduction in the total energy output can be obtained
by inserting a factor ${\cal F}_{r_0} \leq 1$ in
Eq.~(\ref{3.2}), such that
\begin{equation}
E = 2 \kappa {\cal F}_{r_0} \mu^2/M.
\label{3.3}
\end{equation}
An expression for ${\cal F}_{r_0}$ can be found in
Ref.~\cite{Anninos94b}.

The last step consists of invoking the membrane paradigm to calculate
the corrections due to the internal dynamics of the infalling object,
which is now taken to be another black hole instead of a structureless
point particle. These corrections can also be put into the form of
additional factors to be inserted into the expression for $E$.
These factors are all of order unity, implying that the
internal dynamics of the black hole is unimportant in the infalling
process.  so far as gravitational wave generation is concerned, a black
hole falling into another black hole is not much different from that
of a point particle falling into a black hole.
It is found ~\cite{Anninos94b} that the largest effect comes
{}from
the tidal heating of the black hole horizon, which is distorted by the
gravitational field of the other hole as they approach each other.
The corresponding factor ${\cal F}_h$ is found to be
a function of $r_0$, with its
smallest value $\sim 0.86$ realized
when $r_0 \rightarrow \infty$.  For the
range of $r_0$ covered by the numerical evolutions described in
Sec.~II, the total energy
\begin{equation}
E = 2 \kappa {\cal F}_{r_0} {\cal F}_h \mu^2/M,
\label{ef}
\end{equation}
is reduced just by a few percent by ${\cal F}_h$.  (There are other
factors with less than one
percent effects.  They are not considered in this paper.
For details see Ref.~\cite{Anninos94b})

\begin{figure}
\epsfxsize=200pt \epsfbox{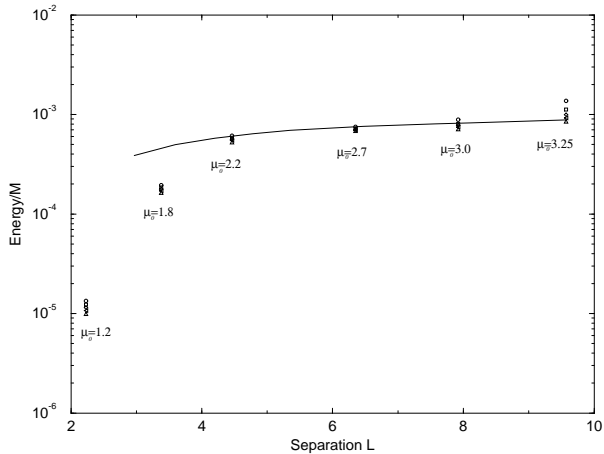}
\caption{Comparison of the particle-membrane approach computation of the
energy vs. the full numerical results.  The solid line gives the
results of the particle-membrane approach, while the clusters of
symbols represent energies computed numerically for a series of
simulations carried out for different values of $\mu_0$.  For each
value of $\mu_0$, the different symbols correspond to energies
extracted at different radii.  For small initial
separations with $\mu_0 \leq 1.8$, with the two holes surrounded by a
common horizon at the initial time,
the particle-membrane approach overestimates the energy output.}
\label{fig3.1}
\end{figure}

We compare the predictions of the particle-membrane approach with
the numerical results in Fig. \ref{fig3.1}. In the figure, the total
energy radiated is plotted as a function of both $L$, the
proper initial separation (a function of $r_0$), and the
Misner parameter $\mu_0$. The clustered symbols show the
energy extracted at various detector locations, as obtained
{}from the numerical results. The procedure for carrying out this
calculation was described in Sec.~II. The solid line is a plot
of Eq.~(\ref{ef}). We see that for sufficiently large initial
separations, the agreement is quite good.
Equation (\ref{ef}) overestimates the energy output when
the initial separation is such that $\mu_0 \leq 1.8$. This
is to be expected, because for such small initial separations,
the two holes are surrounded by a common horizon
\cite{Anninos94f}.  For such cases the particle-membrane
approach, which is based on an object falling into a black hole,
is clearly inappropriate.  Much better suited is
the close-limit approach, to be discussed in Sec.~IV.

\begin{figure}
\epsfxsize=200pt \epsfbox{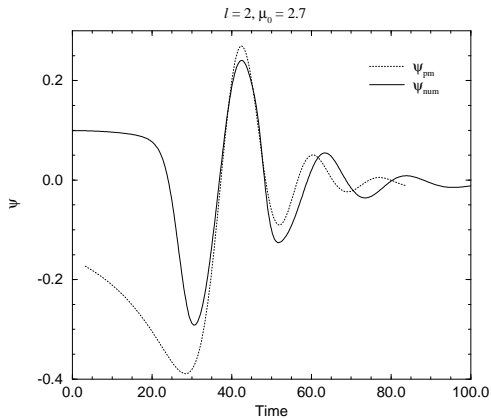}
\caption{Comparison of the particle-membrane approach computation of the
waveform (dotted line) vs. the full numerical results (solid line).
The waveforms were extracted at $r=35M$ and the initial data are for
two black holes separated by a distance $L=6.35M$ ($\mu_0=2.7$).}
\label{fig3.2}
\end{figure}

We now turn to a discussion of the waveforms. In Fig. \ref{fig3.2} we
represent by a solid line the numerically-obtained waveform
$\psi_{\rm num}(t)$, as measured by a detector at $r=35 M$,
resulting from a collision with Misner parameter $\mu_0 =
2.7$ (corresponding to $L=6.35 M$). We wish to compare this
with a waveform obtained using the particle-membrane approach.

To construct a waveform with the particle-membrane approach, we begin
with $\psi_\infty$, the waveform resulting from the infall of a
particle of mass $m_1$ into a black hole of mass $m_2 \gg m_1$, with
the infall starting at infinity. The waveform is measured by a detector
situated near future null infinity; it is a function of retarded time
$t-r^*$,
\begin{equation}
\psi_\infty = \psi_\infty \left(\frac{t-r^*}{m_2}\right),
\label{3.5}
\end{equation}
where $r^* = r+ 2m_2 \ln(r/2m_2-1)$. The function
$\psi_\infty$ has been calculated, using perturbation theory,
by several authors \cite{Detweiler79,Petrich85}.

To convert $\psi_\infty$ into a ``perturbation-membrane'' waveform,
$\psi_{\rm pm}$, we proceed in three steps.  First, we renormalize
the amplitude of $\psi_\infty$ in such a way that the total energy
carried off by that wave agrees with Eq.~(\ref{ef}). Second, we
replace $m_2$ in Eq.~(\ref{3.5}) by $M$, the {\it total} mass of the
spacetime.  This is justified by noting that the waves reaching an
observer situated at a large distance from the system are scattered
(except for some high frequency components) by the curvature of the
spacetime whose length scale is determined mainly by the total mass
$M$ of the system.  [This is different from what was used earlier for
the total energy, which depends on the length scale of the strong
field region experienced by the particle falling into the black hole.
The generation of radiation energy (which strongly depends on near
zone physics)
involves a different scale compared to that of wave propagation
through the field of the combined holes.]  In particular, at late
times the waveform must be dominated by quasinormal ringing of the
{\it final} black hole, which has a mass $m_1+m_2 = M$.  Third, to
compare with the numerical result $\psi_{\rm num}$ which is
calculated for a detector at radius $r=35M$  (instead of at future null
infinity where $\psi_{\rm p}$ is given), we have chosen
the phase of $\psi_{\rm pm}$ so as to match that of $\psi_{\rm
num}$ at late times. The final result is displayed as a dotted line in
Fig. \ref{fig3.2}.

The numerical and particle-membrane waveforms are in reasonably good
agreement when their respective behavior is dominated by quasinormal
ringing.  We notice that at late times, the wavelength of
$\psi_{\rm num}$ becomes longer than that of $\psi_{\rm pm}$.
This is an artifact of the numerical calculation, as was already noted in
Ref.~\cite{Anninos94b}. (The behavior of $\psi_{\rm num}$ at late times is
sensitive to the choice of various parameters, such as
resolution, used in the numerical computation.) There is also a
significant discrepancy at early times: $\psi_{\rm pm}$ displays
a long downward ramp corresponding to the slow infall of the
particle at early times.  Because in the Misner data
the black holes collide from
a finite initial separation, this ramp is not
found in the numerically-obtained waveform.
Indeed, the numerical simulation can cover only cases with
relatively small initial separation, namely, less than $10\, M$.
For larger separations, there are difficulties associated with
the coordinate system used in the numerical calculation, see the
previous
section. We note that
this low frequency downward ramp part of $\psi_\infty$ does not
correspond to much radiation energy, as the luminosity is related
to the time derivative of $\psi$.  We therefore have the good agreement in the
total energy, as shown in Fig. \ref{fig3.1}, despite the differences in the
waveforms.

This discrepancy in the early part of the waveforms suggests
a refined treatment of the particle part of the
particle-membrane approach, to which we now turn.

\subsection{Time-Symmetric Trajectory}
\label{sec:timesym}
\begin{figure}
\vspace{1.5 cm}
\hspace{1cm} \epsfxsize=200pt\epsfbox{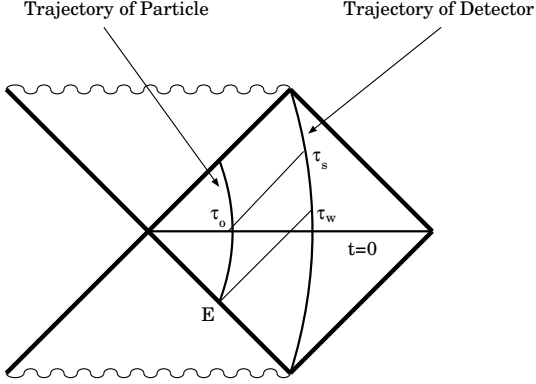}
\vspace{1cm}
\caption{ A time symmetric particle trajectory used in the
membrane-particle approximation model.  The particle emerges from the
past horizon at the point $E$ and falls back in to the future horizon
at a later time.  The times labeled $\tau_w$ and $\tau_s$ are the times
at which signals reach the detector from the events when the particle
leaves the white hole and crosses the point of time symmetry, respectively.}
\label{fig3.3}
\end{figure}

The Misner data describe two Einstein-Rosen throats which first
fly apart from one another, then turn around and fall back in a
time symmetric manner.  The geometry is time symmetric about the
$t=0$ slice at which the Misner data is given [although
the evolution of the black hole horizons are not time symmetric
(Ref.~\cite{science})].
In this subsection we refine our particle-membrane approach by
considering, instead of an infall from infinity, a time-symmetric
trajectory such as depicted in Fig. \ref{fig3.3}. In this
refined model, the particle emerges from the white-hole
(past) horizon, travels up to a radius $r_0$, and then falls
back. The trajectory is a geodesic of the Schwarzschild spacetime,
and is time-symmetric about the turning point $r=r_0$.

This problem not only provides a better model for the collision of
two black holes, it is also interesting in its own right.  We will see
that the particle's emergence from the past horizon causes the
excitation of the {\it white-hole} quasinormal modes.  When the
particle falls back, it is the black-hole quasinormal modes which are
now excited, as we have seen previously.  The white-hole and
black-hole modes are generated with different amplitudes: Although
both the background spacetime and the trajectory are time-symmetric
about $t=0$, the radiation pattern is not.  This should not be
surprising, because the radiation is calculated using {\it retarded}
integrals, which break the time symmetry. To the best of our
knowledge, perturbation-theory calculations of time-symmetric
trajectories have never been carried out before, and the quasinormal
ringing of white holes has never been observed.

To treat this problem we use Teukolsky's perturbation
formalism \cite{Teukolsky73}, as presented in detail in
Ref.~\cite{Poisson}. Schematically, the equation to be solved
is of the form
\begin{equation}
\Box \Psi = \rho,
\label{waveeqn}
\end{equation}
where $\Psi$ represents the perturbation,
$\Box$ a certain wave operator,
and $\rho$ the source term (which is constructed from the
particle's stress-energy tensor). The solution
to Eq.~(\ref{waveeqn}) can be expressed as
\begin{equation}
\Psi = \Box^{-1}_r \rho
\label{wesol}
\end{equation}
with $\Box^{-1}_r$ designating the retarded integral, which
consists of three components: (i) the retarded Green's function,
which is determined by the operator $\Box $, (ii) the source
term $\rho$, and (iii) the initial data on a Cauchy surface.
Our goal is to represent with perturbation
theory a physical situation which is as close as possible to that of
the Misner data.  For (ii), we take the source
term $\rho $ to be given by the
time symmetric trajectory depicted in Fig. \ref{fig3.1}, as discussed above.
For (iii), we shall first study the case of no outgoing wave
{}from the past horizon $r = 2 m_2$ and no incoming wave from
past null infinity $r = \infty$.  We shall return to a more
careful consideration of this point later.

With (ii) and (iii) so specified, it is straightforward to write
down the Zerilli function expressed
for each multipole $\ell$ as the Fourier integral
\begin{equation}
\psi_p(t-r^*) = 2 \sqrt{2} m_1 \int_{-\infty}^\infty d\omega\,
Z_\ell(\omega) e^{-i\omega(t-r^*)},
\label{3.6}
\end{equation}
where
\begin{eqnarray}
Z_\ell&&(\omega) = \frac{i}{\omega A_\ell(\omega)}
\frac{\sqrt{(\ell-1)\ell(\ell+1)(\ell+2)}}{
(\ell-1)\ell(\ell+1)(\ell+2) - 12im_2\omega}\nonumber \\
&&\,\times
\sqrt{\frac{2\ell+1}{4\pi}}
\sqrt{\frac{r_0}{2m_2}}
\int_{-\eta_H}^{\eta_H} d\eta\,
\left( \frac{\sqrt{1-2m_2/r_0} + \dot{r}}{1-2m_2/r} \right)^2
 \nonumber\\
&&\times e^{i\omega t}\,\Gamma_{\ell}(\omega) X_\ell (\omega;r).
\label{3.7}
\end{eqnarray}
Here, $\eta$ is a parameter
along the trajectory, such that $r= (r_0/2) (1+
\cos\eta)$, $\dot{r} \equiv dr/d\tau = - (2m_2/r_0)^{1/2}
(1+\cos\eta)^{-1}\sin\eta$, with $\tau$ denoting proper time,
and
\begin{eqnarray}
t =&& \frac{1}{2} r_0
\sqrt{\frac{r_0}{2m_2}-1}
\left[ \left(1+\frac{4m_2}{r_0} \right) \eta
+ \sin\eta \right] \nonumber\\
&&+ 2m_2 \ln \left|
\frac{\sqrt{r_0/2m_2-1}+\tan \frac{1}{2} \eta}{
\sqrt{r_0/2m_2-1}-\tan \frac{1}{2} \eta} \right|;
\label{3.8}
\end{eqnarray}
$\eta$ increases monotonically along the trajectory, and takes
the values $\mp \eta_H$, where $\eta_H = \cos^{-1}(4m_2/r_0-1)$,
at the past and future horizons, respectively. The
function $X_\ell(\omega;r)$ satisfies the Regge-Wheeler equation,
\begin{equation}
\Biggl\{ \frac{d^2}{dr^{*2}} + \omega^2\! - \!
\Biggl(1\!-\! \frac{2m_2}{r}\Biggr)
\Biggl[\frac{\ell(\ell-1)}{r^2}\! -\! \frac{6m_2}{r^3} \Biggr]
\Biggr\} X_\ell\! =\! 0,
\label{3.9}
\end{equation}
with boundary condition $X_\ell(\omega;r\to 2m_2) \sim
e^{-i \omega r^*}$. At large values of $r$, the Regge-Wheeler
function becomes $X_\ell(\omega;r\to\infty) \sim A_\ell (\omega)
e^{-i\omega r^*} + O(e^{i\omega r^*})$, which defines
the constant $A_\ell (\omega)$ appearing in Eq.~(\ref{3.7}).
Finally, $\Gamma_\ell(\omega)$ is the differential
operator
\begin{eqnarray}
\Gamma_\ell(\omega) &=&
2(1-2m_2/r+i\omega r) r (1-2m_2/r) \frac{d}{dr}  \nonumber\\
& &+(1-2m_2/r) \left[ \ell(\ell+1) - 6m_2/r \right]
\nonumber \\
& & \mbox{} + 2 i \omega r (1-2m_2/r+i\omega r).
\label{3.10}
\end{eqnarray}

\begin{figure}
\epsfxsize=200pt \epsfbox{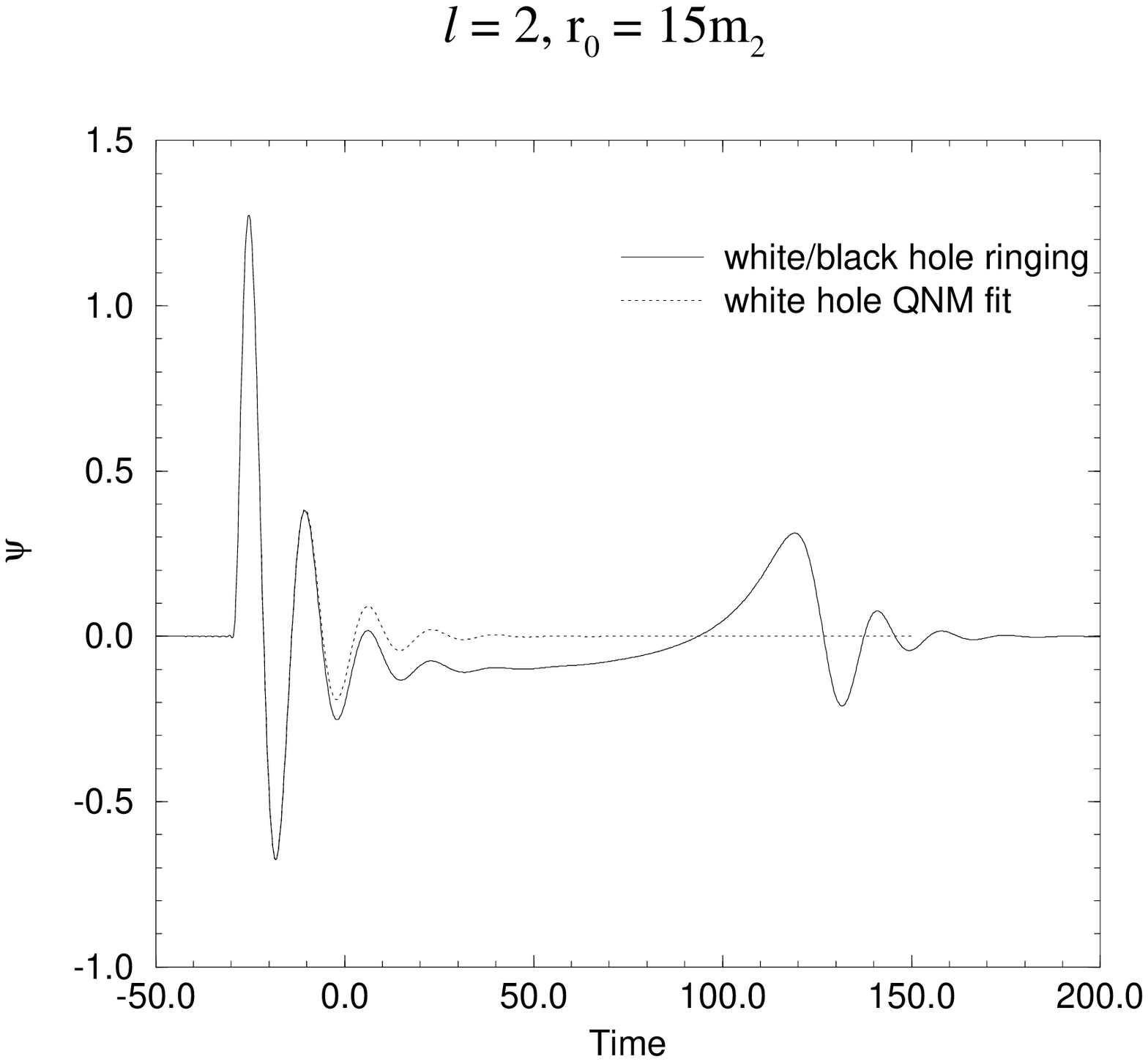}
\epsfxsize=200pt \epsfbox{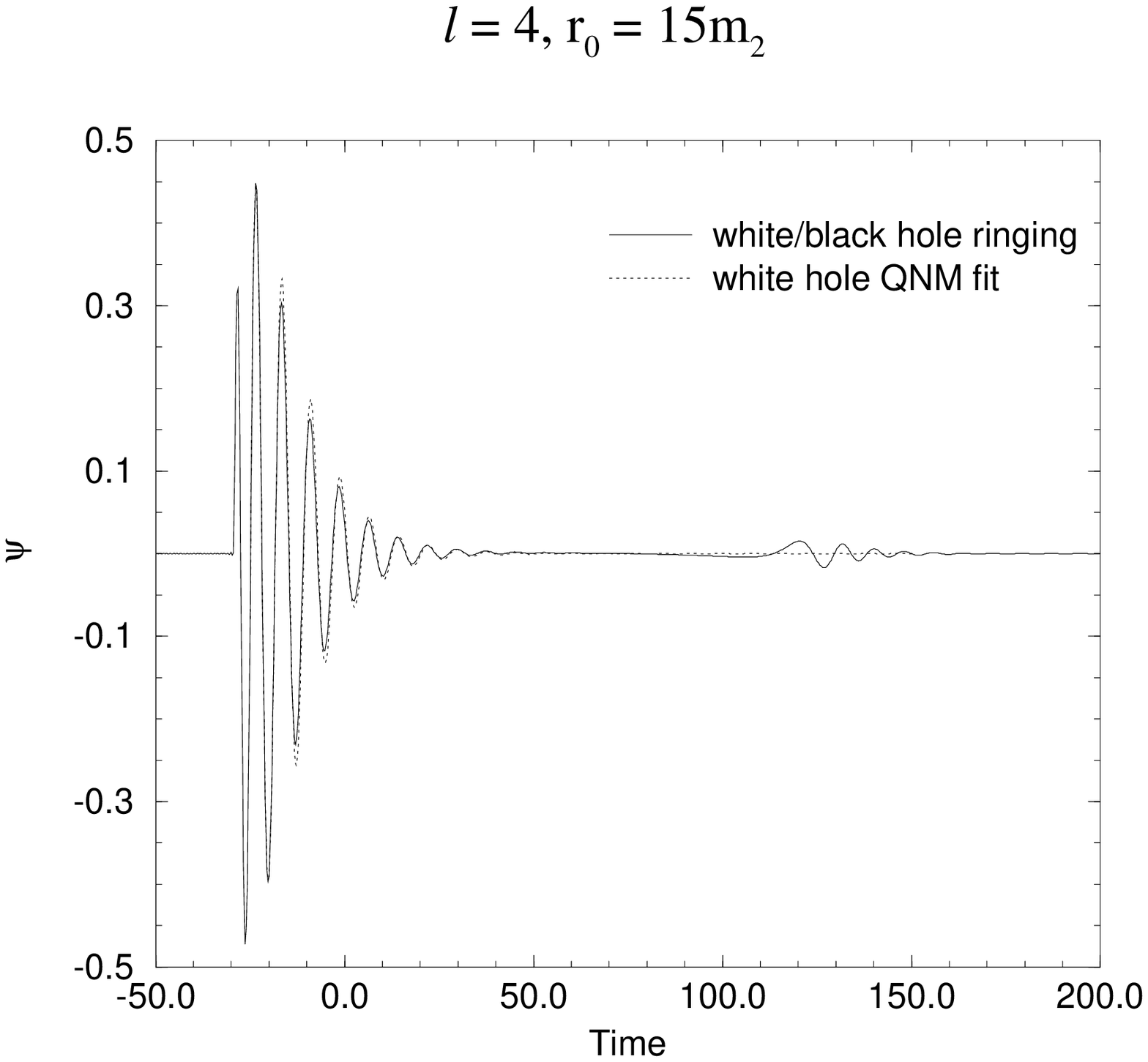}
\caption{The $\ell=2$ and $\ell=4$ components of $\psi_p$ for the case
$r_0=15m_2$; the Zerilli function is plotted as a function of
coordinate time $t$, and is measured by a detector situated at radius
$r=40m_2$.  There are three components in the waveform: white hole
ringing, direct bremsstrahlung radiation, and black hole ringing.
The dotted lines show the quasinormal mode fit to the
white
hole ringings.}
\label{fig4}
\end{figure}

In Fig. \ref{fig4}a we show the $\ell=2$ component of $\psi_p$ for the
case $r_0=15m_2$; the Zerilli function is plotted as a function of
coordinate time $t$, and is measured by a detector situated at radius
$r=40m_2$. At $t=-29.5m_2$ a sharp feature appears, corresponding to
the particle emerging from the past horizon ($\tau_w$ in
{}~Fig. \ref{fig3.3}). The white hole subsequently goes into quasinormal
ringing. In Fig.~\ref{fig4}a the dotted line represents a pure
quasinormal-mode signal, the superposition of the first two (the least
damped) $\ell=2$ quasinormal modes of a Schwarzschild spacetime of
mass $m_2$, with amplitude and phase determined by matching to the
white-hole ringing. The dotted line was plotted starting from $t=-22
m_2$, but it is practically indistinguishable from the solid line in
the early part.  The good agreement between the wavelengths confirms
that the early-time portion of $\psi_p$ is indeed quasinormal
ringing. The ringing eventually stops, and the subsequent
low-frequency signal at $30 m_2 <t< 80 m_2$ is the bremsstrahlung
radiation coming directly from the particle as it emerges from the
strong field region. As the particle moves outside $r = 3 m_2$, which
is the peak of the potential barrier for the scattering of
gravitational waves, the bremsstrahlung radiation can reach the
observer without much scattering by the spacetime curvature.  The ray
emitted at the turning point reaches the detector at time $t=27m_2$
(cf. point $\tau_s $ in ~Fig. \ref{fig3.3}).  The particle then turns
around and falls back.  As the particle once again goes through the
strong field region while approaching the future horizon of the black hole,
quasinormal modes are excited once more. It can be seen from the figure
that the black-hole ringing has a smaller amplitude than the white
hole's.

In Fig.~\ref{fig4}b, we show the $\ell=4$ component of $\psi_p$ for
the same case.  We see that the radiation is again consist of the same
three components, except that the bremsstrahlung contribution to
$\ell=4$ is much weaker.  The quasinormal mode nature in the early
part of the waveform is clear, with the dotted line giving the
quasinormal mode fit (first two $\ell=4$ modes).  The dotted line
begins at $t=-25 m_2 $ and is nearly indistinguishable from the actual
waveform.  Notice that the black hole ringing has a much lower amplitude.

The waveforms given in Figs.~\ref{fig4}ab point to the fact that the
Misner data represent more than just two black holes in time symmetric
trajectories.  If there were no wave coming from the past null
infinity in the Misner data, we expect that there would be ``white
hole ringing'' as shown in Figs.~\ref{fig4}ab, for black holes in time
symmetric motion ~\cite{science}.  As the Misner data produces no
white-hole ringing for $t > 0$, independent of the location of the
detector, the initial data given at $t=0$ must contain the right
amount of waves traveling inward to cancel the out-going ``white hole
ringing''.  The time symmetry implies that there is no net flux at
$t=0$.  The fact that there must be waves coming in from past null
infinity is also guaranteed by time symmetry: there are as many waves
coming in from past null infinity as going out to future null
infinity.  The same argument can also be applied to waves in and out
of future and past horizons.

These considerations suggest that, for a meaningful comparison to the
Misner data, the perturbation calculation we should carry out is {\it
not} exactly the $\psi_p$ given by Eq.~(\ref{3.6}) above, which
assumes no out going waves from the past horizon ( $ \it H ^ - $) and
no in coming waves from
past null infinity ( $I^- $).  Instead, nontrivial initial
data [component (iii) in the specification of the retarded integral,
Eq.~(\ref{wesol})] should be imposed on $ \it H ^ - $ and $
I^ - $ to ensure that there is no net flux at $t=0$.  In
particular, the Cauchy development of these data should cancel the
white hole ringing.

\begin{figure}
\epsfxsize=200pt \epsfbox{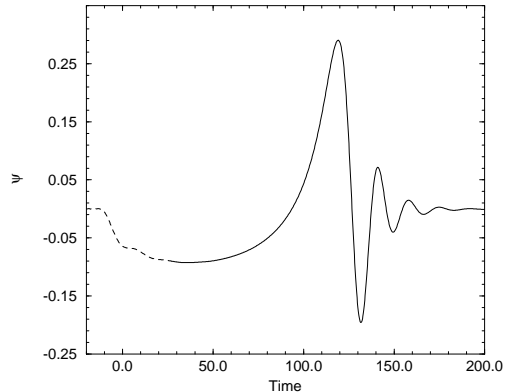}
\caption{We show the $\ell=2$ component of the particle-membrane
waveform for the case $r_0 = 15 m_2$.
The present numerical relativity code cannot evolve initial
data sets with such large separations long enough to determine
the full waveform.}
\label{fig5}
\end{figure}

Obviously, the construction of such data on $ \it H ^ - $ and $ I^ - $
is difficult, if at all practical.  We circumvent this difficulty by
directly subtracting the ``offending'' components of $\psi_p$ shown in
Figs. \ref{fig4}.  [What this subtraction corresponds to at $ \it H ^
- $ and $ I^ - $ can in principle be determined by integrating
backward in time.  But that is not the concern of this paper.]  The
result of this subtraction is shown in figure \ref{fig5}. The
components that we want to subtract can readily be identified: First,
the white hole ringing.  Second, we note that the portion of the wave
given as a dashed line has retarded time less than 0.  It is emitted
by the particle as it is flying out before reaching $r_0$ ($t < \tau_s
$ in Fig.
\ref{fig3.3}), and therefore
has no corresponding part in the Misner data.  [As the scattering due
to the potential is weak once the particle is outside the peak of the
potential barrier at $r = 3 m_2$, this contribution to the waveform can
be identified by its retarded time.  This is {\it not} so for quasinormal
ringing, which is a multiple scattering phenomenon.]

These subtractions based on physical understanding of the system,
although not expected to yield mathematically exact time
symmetric spacetimes, make it possible to use perturbation
theory to construct a waveform
which is comparable to that computed numerically.
Both its similarities and differences with the
numerical results shed light on the physical meaning of the Misner
data.

The next step in the construction of the waveform is to extrapolate
our results to the equal-mass case.  One reasonable choice is to
replace $m_1$ with $\mu$ (the reduced mass), while all $m_2$ in
Eqs.~(\ref{3.6})--(\ref{3.10}) with $M=m_1+m_2=2m_2$ (the total mass).
This respectively fixes the amplitude of the Zerilli function, and the
units of the time coordinate $t$.  We also correct for the internal
dynamics of the ``particle'' (now imagined to be a black hole) by
multiplying the Zerilli function by $({\cal F}_h)^{1/2}$ [c.f.
Eqs.~(\ref{ef})]. The result is our particle-membrane waveform,
$\psi_{\rm pm}$, displayed in Fig. \ref{fig5} for the $\ell =2$
component.  This figure corresponds to the
case where the maximum separation between the two black hole is
$r_0=15m_2$.

It was useful to consider the case $r_0=15m_2$ because the
gravitational-wave signal shows (before subtraction) three
relatively well separated
portions: white-hole ringing, particle bremsstrahlung, and
black-hole ringing. However, and unfortunately, numerical
results are not available for such a large initial separation.
(Difficulties associated with the \v{C}ade\v{z} coordinate
system make the numerical evolution unreliable when the initial
separation is large).
So Fig. \ref{fig5} can be regarded as
a ``prediction'' of what the numerical evolution of the Misner
initial data should
produce when pushed to such initial separations.

\begin{figure}
\epsfxsize=200pt \epsfbox{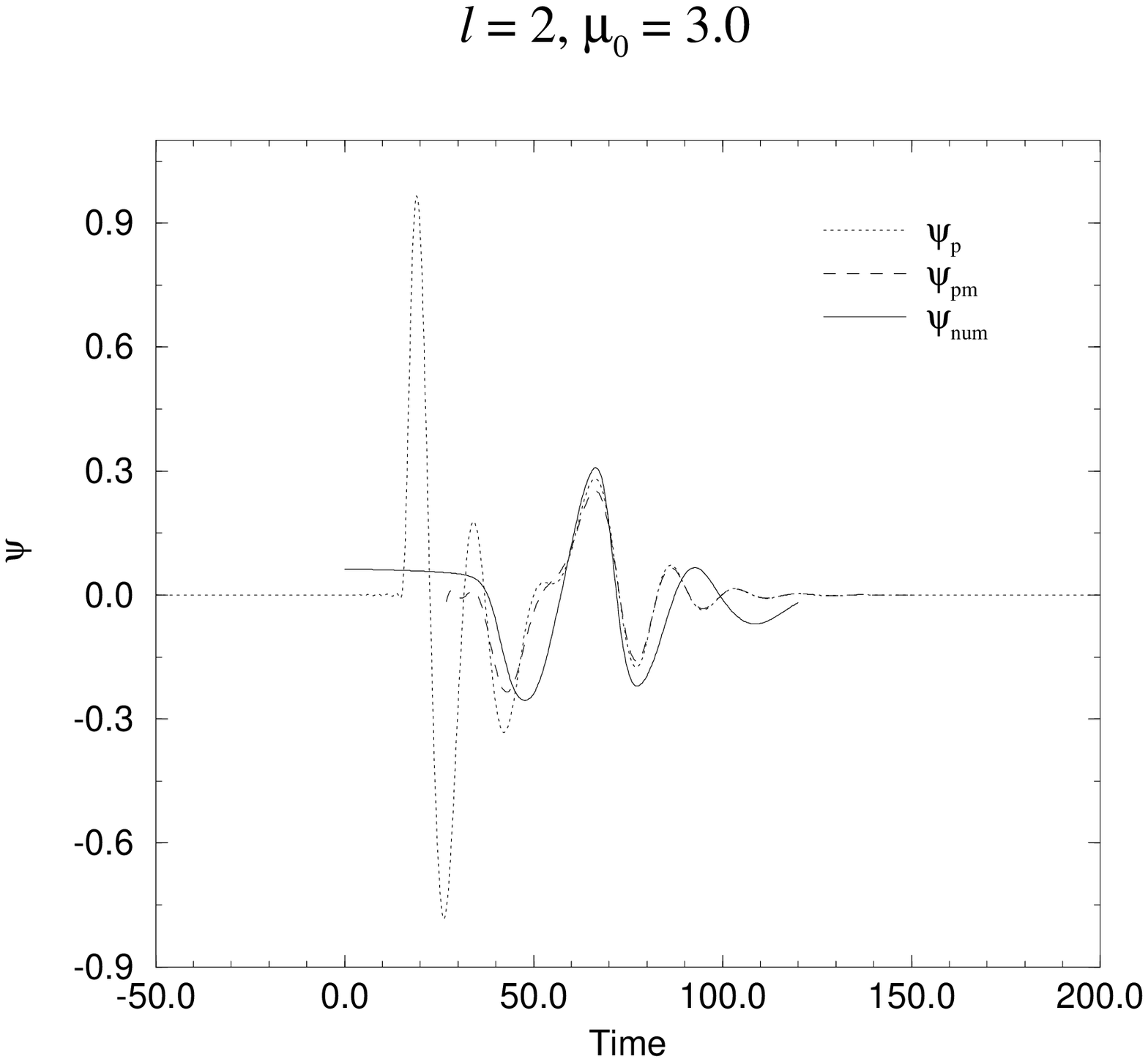}
\epsfxsize=200pt \epsfbox{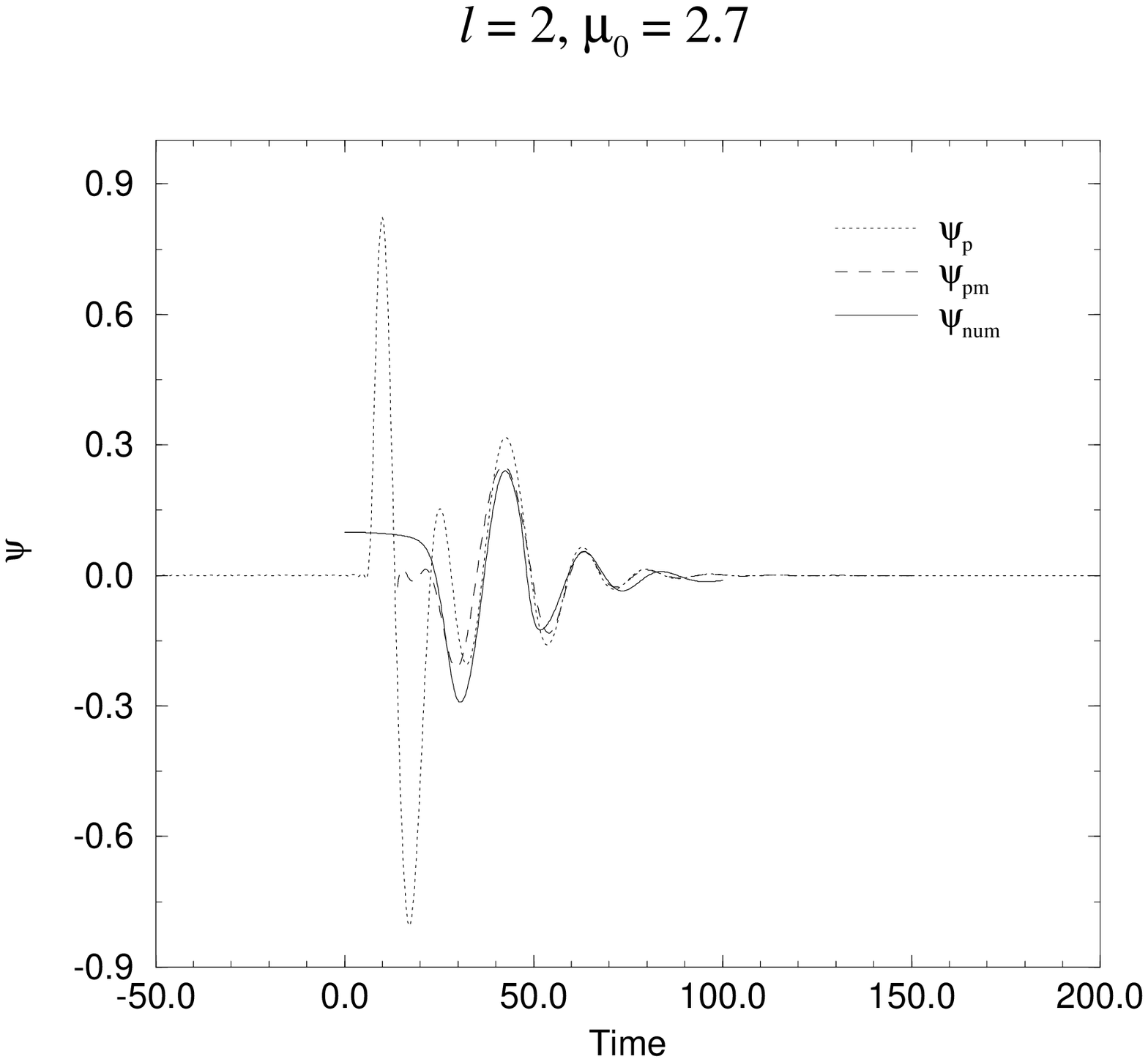}
\caption{Comparison of the $\ell=2$
particle-membrane waveforms with the numerical results for separations
$\mu_0=3.0$ and $\mu_0=2.7$.  The dotted line is the result of the
particle trajectory before the ``white hole'' part has been
subtracted, the dashed line is the particle-membrane waveform
corrected for the white hole ringing, and the solid line is the full
numerical result.  }
\label{fig6}
\end{figure}

To compare $\psi_{\rm pm}$ to existing numerical waveforms, we
consider the cases $r_0 = 6.8M$ (corresponding to a value $\mu_0 =
3.0$ for the Misner parameter) and $r_0 = 5.4 M$ ($\mu_0 = 2.7$).  For
such initial separations, the three portions of the waveforms are not
cleanly separated. In Fig.~\ref{fig6}a we consider the case $\mu_0=3.0$, and
display the $\ell=2$ component of three distinct waveforms; the
waveforms are measured by a detector situated at $r=35M$. The first is
$\psi_p$ represented by the dotted line, obtained from Eq.~(\ref{3.6})
(with extrapolation $m_1 = \mu$,$ m_2 = M$ ). The second is $\psi_{\rm
pm}$ represented by the dashed line, obtained from $\psi_p$ by
subtracting the white-hole ringing and including the corrections due
to internal dynamics. The third is the numerical waveform, $\psi_{\rm
num}$ as solid line. In Fig.~\ref{fig6}b we do the same for the case
$\mu_0=2.7$ (the detector is situated at $r=20M$).

In Fig.~\ref{fig6}a, the white-hole ringing produces peaks at around $t/M =
(19,35,52)$; the last peak overlaps with the black-hole ringing.  The
bremsstrahlung radiation is also contained in the overlap.  Because of
the overlap with the bremsstrahlung radiation, the white-hole ringing
cannot be subtracted efficiently, and the resulting $\psi_{\rm pm}$
remains contaminated. The contamination is less in Fig.~\ref{fig6}b, because
less bremsstrahlung is present in the signal (the magnitude of the
bremsstrahlung radiation is determined by the speed of the particle in
the part of the trajectory with $r> 3 M$).  Moreover, in this case the
white-hole ringing is more in phase with the black hole's, hence the
contamination less noticeable.  We do not see any ``hump'' like
the one that appeared at $t=52 M$ in Fig. 6a.  Instead, the contamination
showed up as a slight lengthening of the wavelength of $\psi_{\rm
pm}$.  Apart from these effects, the agreement between $\psi_{\rm pm}$
and $\psi_{\rm num}$ is reasonably good.

The flat plateau at early times ($t=0$ to 37 for $\mu =3.0$ and
$t=0$ to 25 for $\mu =2.7$) in $\psi_{\rm num}$ is a non-radiative
component in the Misner data.  It's magnitude depends on the location
of the detector, chosen to be located at $35\, M$ for
$\mu = 3.0$ and $20\, M$ for $\mu =2.7$, with the former one having
a much smaller amplitude.  Should the detector be
put further out, this component would be even smaller.  After the
quasi-mode ringing sets in, we see that $\psi_{\rm num}$ agrees
very well with $\psi_{\rm pm}$ both in their phases and amplitude.
This is remarkable, in view of the fact that the
matching involves no adjustable parameters.

\begin{figure}
\epsfxsize=200pt \epsfbox{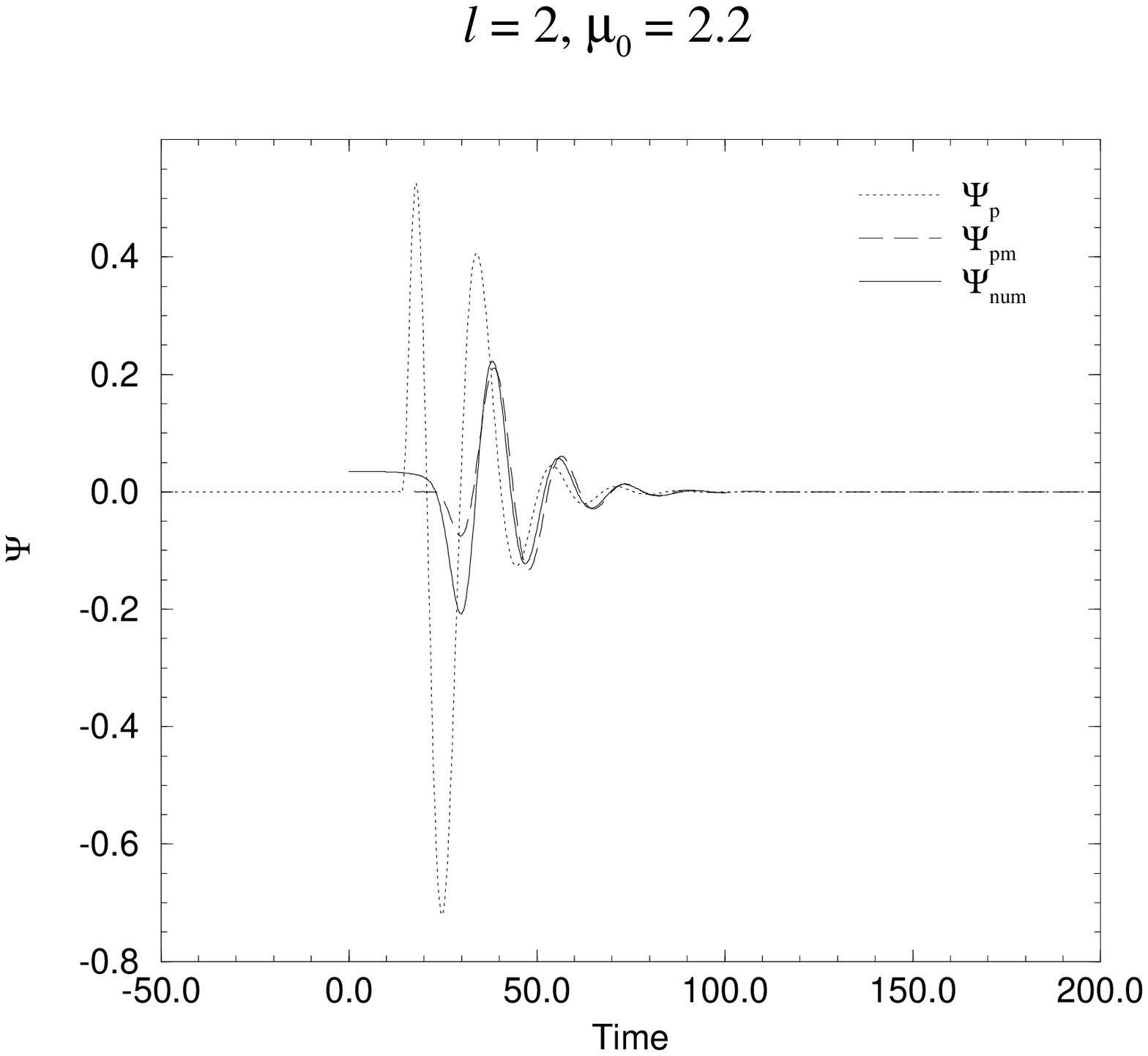}
\epsfxsize=200pt \epsfbox{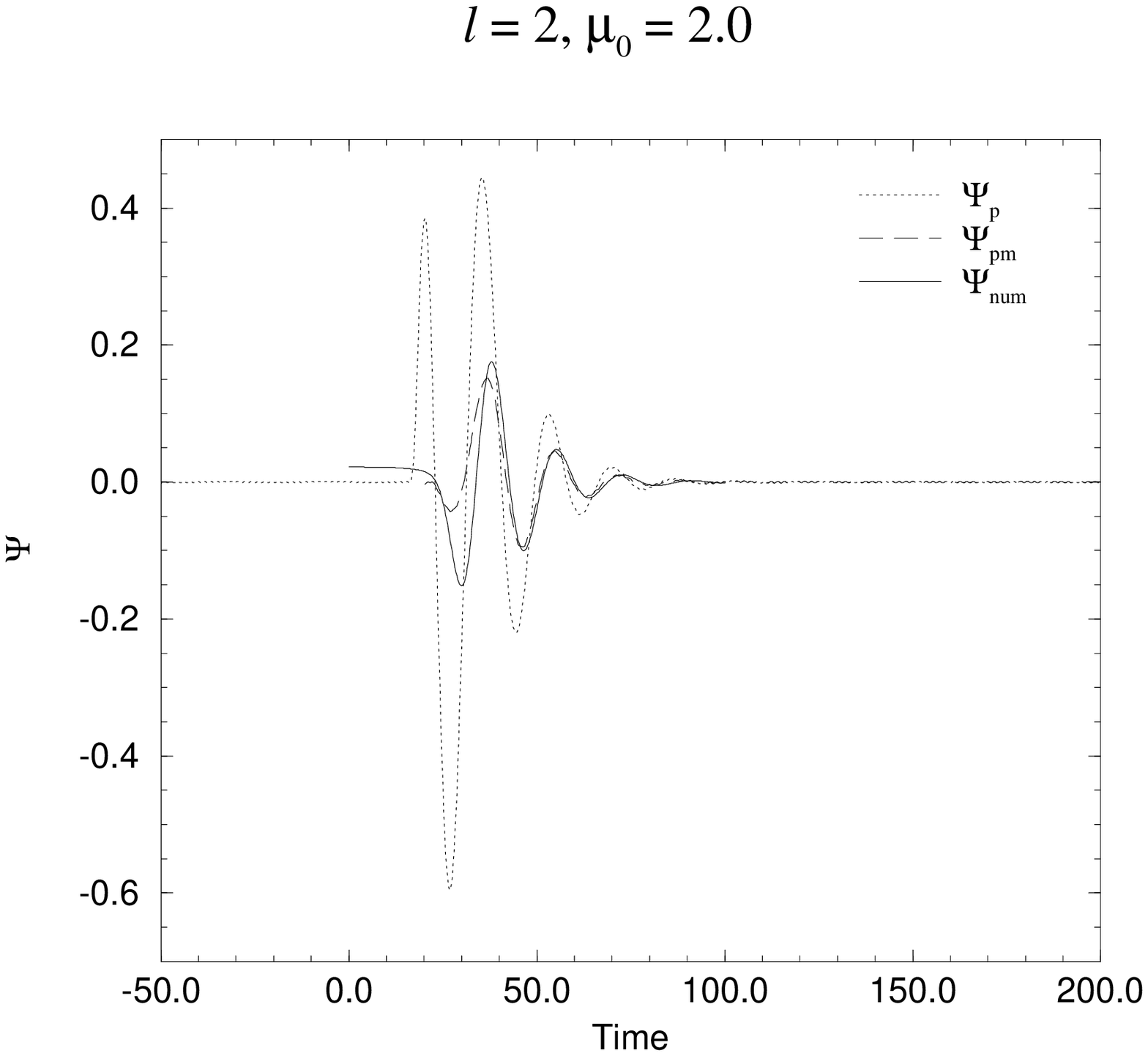}
\caption{Comparison of the $\ell=2$
particle-membrane waveforms with the numerical
results for separations $\mu_0=2.2$ and 2.0.}
\label{fig7}
\end{figure}

In Figs.~\ref{fig7}a and \ref{fig7}b we push the particle-membrane
approach to even smaller initial separations. In Fig.~\ref{fig7}a we
display the waveforms corresponding to the case $\mu_0=2.2$ ($r_0 =
3.7M$), while in Fig.~\ref{fig7}b we consider the case $\mu_0=2.0$
($r_0=3.2 M$). We see that, for such initial separations, the waveform
$\psi_p$ obtained from the perturbation calculations (dotted lines in
Figs.~\ref{fig7}a and \ref{fig7}b ) has its white-hole and black-hole
ringing parts merged to a large extent. This is expected, because
$r_0$ is very close to $3M$, the location of the peak of the potential
barrier surrounding the hole. For such cases, in order to separate out
the white hole ringing, we match $\psi_p$ to {\it two} sets of
quasinormal modes which are excited at two different times.  The
first set is denoted $\psi_{\rm w}$ (``w'' stands for ``white''), and
is composed of the first two $\ell=2$ quasinormal modes.  The second
set which sets in at a later time is denoted $\psi_{\rm b}$ (with
``b'' standing for ``black''), has the same frequencies. With the
amplitude of $\psi_{\rm w}$ determined through such a matching, we
obtain $\psi_{\rm pm}$ by subtracting $\psi_{\rm w}$ from $\psi_p$.
In Figs.~\ref{fig7}a and \ref{fig7}b, $\psi_{\rm pm}$ is represented
as dashed lines.  The agreement between $\psi_{\rm pm}$ and $\simeq
\psi_{\rm num}$ is quite satisfactory except for the initial pulse.
The disagreement in the initial pulse is due to the difficulty in
determining $\psi_{\rm w}$ from the merged white hole-black hole
ringing.  For the case of $\mu_0=2.0$, we find that the
the white hole ringing
$\psi_{\rm w}$, which is to be subtracted away, is quite sensitive to the
fine details in the form of $\psi_p$.  This
signals the beginning of the breakdown of the method.
As pointed out above, we do not
expect our particle-membrane approach to be applicable for
small initial separation between the two holes.  For initial data with
$\mu_0=1.8$ or smaller, the two holes are surrounded by a common event
horizon.  This breaking down of the method at around $\mu_0=2.0$
is consistent with the results on energy radiated
as shown in Fig. \ref{fig3.1}.  The particle-membrane result also begins to
deviate from the numerical result at around $\mu_0=2.0$ there.

\begin{figure}
\epsfxsize=200pt \epsfbox{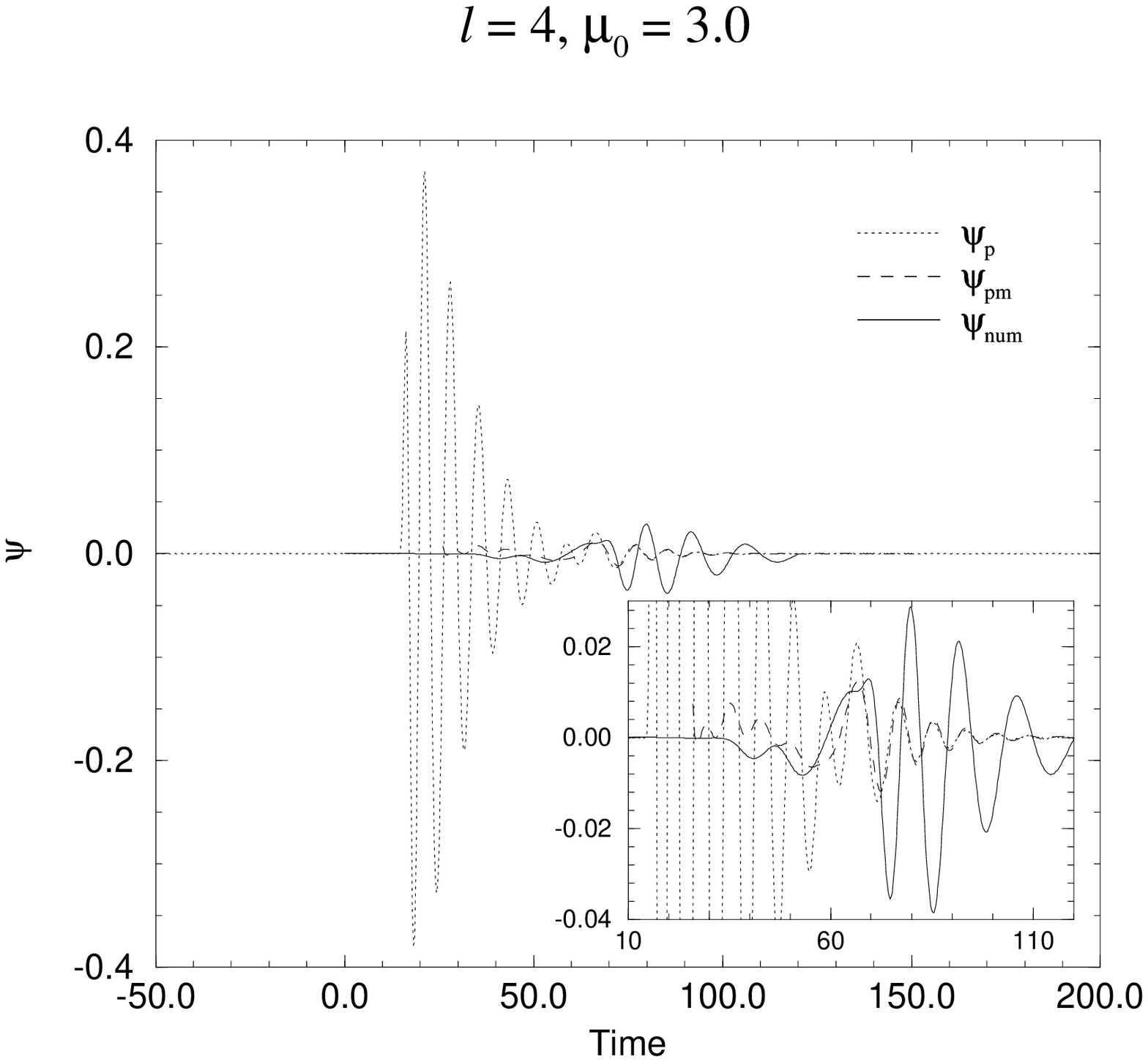}
\epsfxsize=200pt \epsfbox{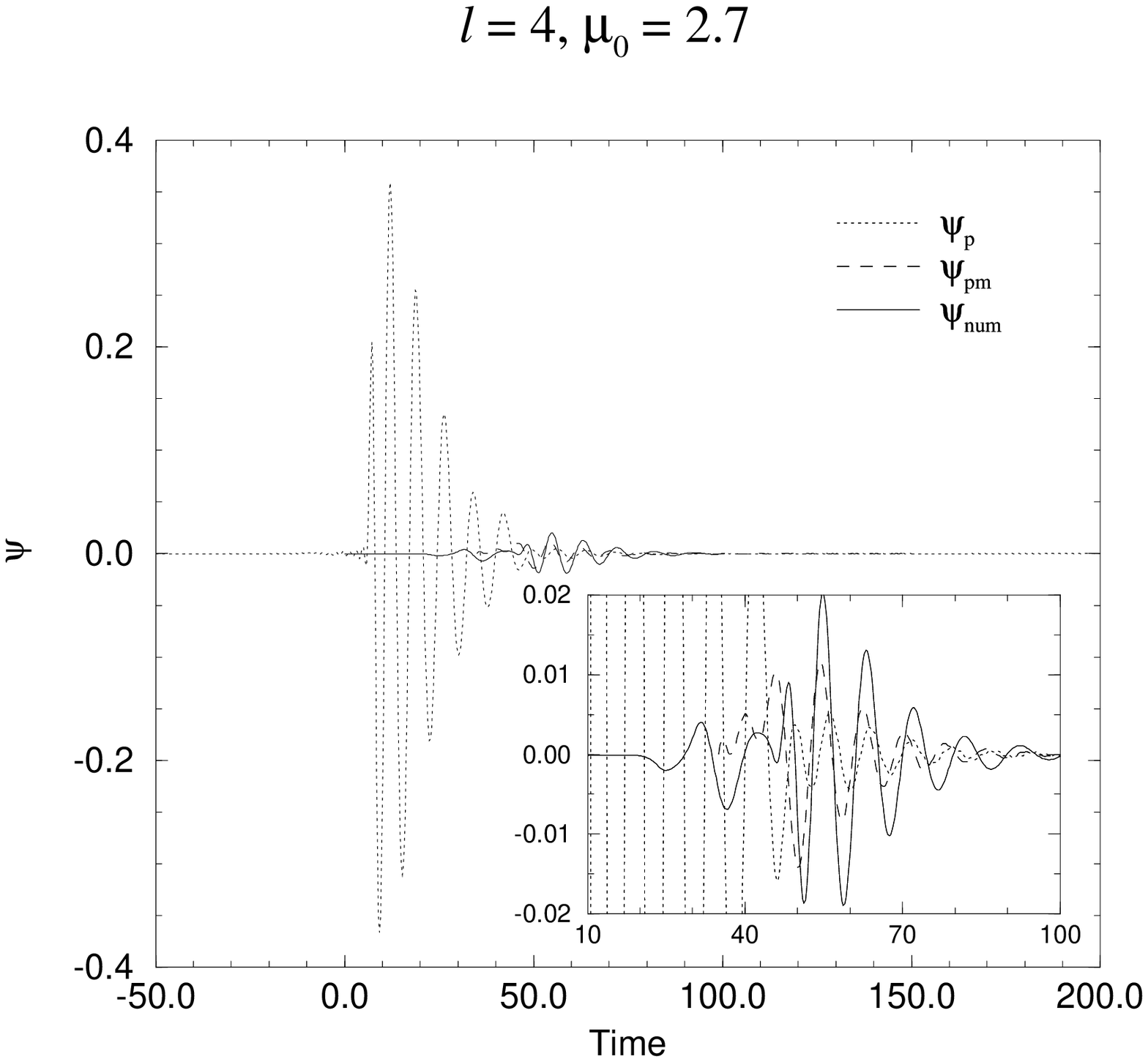}
\caption{Comparison of the $\ell=4$
particle-membrane waveforms with the numerical
results for separations $\mu_0=2.7,3.0$.}
\label{fig8}
\end{figure}

In Figs.~\ref{fig8}, we compare the $\ell=4$ components of the
waveforms for various cases.  The dotted lines give $\psi_p$.  The
dashed lines give the particle-membrane waveform $\psi_{\rm pm}$
extracted from $\psi_p$.  They are superimposed on the numerical
results (solid lines).  Fig.~\ref{fig8}a is for the case $\mu_0=3.0$, while
Fig.~\ref{fig8}b is for $\mu_0=2.7$.  We see that the white-hole
ringing has much larger amplitude than the black-hole ringing (which are shown
blown up in the insets), making the
extraction more difficult.  However, we believe that the $\psi_{\rm
pm}$
so obtained is still more reliable than the numerical ones, which are
considerably larger in amplitude.  In the next section we shall see a
similar situation for small $\mu$ values under the ``close
approximation''.  Although the numerical calculations give waveforms
agreeing very accurately with the semi-analytic methods for the
$\ell=2$ components, the agreement for the $\ell=4$ components is
much less satisfactory.

We conclude this subsection with the following remarks.
In view of the complications associated with the removal of the
white-hole ringing from the perturbative waveform, it is tempting to
rule that only the infall part of the trajectory should be included as
source from the first place.  Thus, one might ignore the first half of
the motion, in which the particle travels from the past horizon to the
turning point at $r=r_0$.  One might then modify Eq.~(\ref{3.7})
so that the integral is now evaluated between $\eta = 0$ and $\eta =
\eta_H$. However, such a truncation would produce nonsense: It
corresponds to the sudden creation of a particle at $r=r_0$.  Such a
violation of the Einstein's equations, which guarantee the
conservation of energy-momentum, produces unacceptable results even at
the linearized level considered here. This has also been observed
previously in Ref.~\cite{kip}.

We should note that our present calculation, as given by
Eqs.~(\ref{3.6})--(\ref{3.10}), has also violated the Einstein
equations.  The Teukolsky equation was integrated assuming no outgoing
radiation from the entire past horizon.  However, this is not possible
with a particle flying out of it ({\it cf.,} Fig.~3).  If one imposes
a no out going wave condition as initial data for the portion of the
horizon at times later than the emergence point of the particle (point
$E$ in Fig.~3), then we cannot impose the same on the portion of the
horizon before point $E$.  Otherwise, Einstein's equations are
violated at the point $E$, with the source term $\rho$ being a
$\delta$ function having support there.  To satisfy Einstein's
equations, one has two options.  One can let the particle trajectory
extend all the way back to the initial singularity, eliminating the
need of specifying boundary conditions on the past horizon.  While
allowing the passage through the coordinate singularity at $r=2m_2$
should be straightforward, the difficulty is that there is no a priori
clear way to pick the initial data at the past singularity that best
corresponds to the Misner data.  Another option is, while imposing a
no out going radiation condition for the portion of the horizon at
times later than the emergence point $E$ of the particle, we integrate
the Teukolsky equation backward along the past horizon across the
point $E$ to determine the suitable data on the earlier part of the
horizon.  However, we note that doing this extra work to get a more
``correct'' set of initial data will not affect much our final
waveform $\psi_{\rm pm}$.  This is because for the data so determined
on the early part of the past horizon, the high frequency components
of it will propagate out unscattered, reaching the detector at a very
early retarded time and hence cannot not interfere with the waveform
we want to extract, while the low frequency components of it will be
trapped by the potential barrier and appear later as quasinormal
ringing of the white hole, which we subtract away anyway.  There will
only be a negligibly small part multi-scattered by the potential
outside and away from its peak $(r \gtrsim 3\, M)$ that can reach the
detector at the time of our interest.  Hence we see that, for our
present purpose, carrying out this extra step of imposing correct data
on the early part of the past horizon can have only negligible effect
on our final waveform $\psi_{\rm pm}$.  This also explains why
violating the Einstein equations in our present setup is not
troublesome (whereas it {\it is} troublesome if the trajectory is
truncated at the turning point), as evident in the close matching of
our results to the numerical results shown above.

To sum up, in this section we have calculated the waveform generated
by a particle in time symmetric motion, with no wave coming from
the past horizon and the past null infinity.  We then subtract
away the
part of the radiation emitted by the particle {\it before} the time
symmetric point $t=0$.  The resulting waveform, with a
correction factor (of order unity) put in to correct for
the internal dynamics of a black hole, is compared to the
waveform obtained by numerically evolving the Misner data.  We
find that for $\mu > 2.0$, the agreements in the phases, frequencies
and amplitudes of the two waveforms are satisfactory.  This
implies that the Misner data represents, to a good approximation,
such a physical situation, namely, two throats in time
symmetric motion
with radiation from the $t<0$ part of the trajectories balanced
by waves from the past horizon and the past null infinity.
For cases with smaller
$mu$ values, we have difficulty in identifying, and hence
subtracting the radiation emitted by the ``particle''
before the time symmetric point, and we cannot obtain a waveform
{}from this semi-analytic approach.  Fortunately, for smaller
$mu$ values, we have developed a different semi-analytic treatment
which we now turn to.

\section{Perturbation Theory for The Close Limit}
\label{sec:pert}
\subsection{Formulation}

When black holes start sufficiently close to each other the analysis
of the radiation generation can be considerably simplified.  From
numerical evolution computations it is known that for $\mu_0$ less
than $\sim1.8$ a common horizon initially surrounds both throats
of the Misner geometry \cite{Anninos94f}. We describe here how the
spacetime exterior to this horizon can be viewed as a distorted
Schwarzschild geometry and can be treated with methods from
perturbation theory.

Our starting point for the perturbation analysis is the Misner
geometry\cite{Misner60},
\begin{equation}
  ds^{2}_{\rm Misner} = a^{2}\varphi^{4}_{\rm Misner}
\left[d\mu^{2}+d\eta^{2} +\sin^2{\eta}\,d\phi^{2}\right]\,,
\end{equation}
where
\begin{equation}
   \varphi_{\rm Misner}
= \sum_{n=-\infty}^{n=-\infty}
  \frac{1}{\sqrt{\cosh(\mu+2n\mu_0)-\cos\eta}}\,,
\end{equation}
and where the coordinates have the range
\begin{equation}
  -\mu_0\leq\mu\leq\mu_0\, \qquad
   0\leq\eta\leq\pi\, \qquad
   0\leq\phi\leq2\pi\quad .
\end{equation}
This geometry represents an asymptotically flat  three geometry
with ADM mass
\begin{equation}\label{Mass}
   M = 4a\sum_{n=1}^{\infty}
       \frac{1}{\sinh{n\mu_{0}}}\equiv4a\Sigma_1 \quad .
\end{equation}
The geometry has two ``throats'' which end at $\mu=\pm\mu_0$.
These throats can be considered to be joined to form a wormhole,
as in the original Misner paper, in which case the Misner solution
is viewed as periodic in $\mu$, with period $2\mu_0$.
Alternatively both throats can be extended into a second
asymptotically flat space\cite{Misner63}, in which case
the period is $4\mu_0$ and range $\mu_0\leq\mu\leq3\mu_0$
may be considered to describe the mirror image solution
with the second asymptotically flat space.

The nature of the Misner metric as a single perturbed hole becomes
clearer if new coordinates $R,\theta$ are introduced as if
$\mu, \eta$ were bispherical coordinates being transformed
to spherical polars:
\begin{equation}\label{transforms}
   R \equiv a
 \sqrt{\frac{\cosh{\mu}+\cos{\eta}}{\cosh{\mu}-\cos{\eta}}}\qquad
   \tan{\theta}\equiv\frac{\sin{\eta}}{\sinh{\mu}}\quad .
\end{equation}
If we want to restrict our $R,\theta$ coordinates to the region
$-\mu_0\leq\mu\leq\mu_0$ we must in principle remove the interiors of
two ``circles'' corresponding to $|\mu|>\mu_0$.

In terms of these new coordinates, the Misner geometry
takes the form
\begin{equation}
  ds^2_{\rm Misner} = \Phi^4(R, \theta; \mu_0)
\left(dR^2+R^2\left[d\theta^2 +\sin^2\theta\,d\phi^2\right]\right)
\end{equation}
with the  conformal factor $\Phi$ given by,
\begin{eqnarray}
  \Phi = 1 +\sum_{n \neq0}&&
  \sqrt{\cosh{\mu}-\cos{\eta}} \times\\
&&\left\{2\sinh^2{n\mu_0}+\sinh{\mu}\sinh{2n\mu_0}\right.\nonumber\\
&&\left. +(\cosh{\mu}-1)
   \cosh{2n\mu_0}+1-\cos{\eta}\right\}^{-1/2}.\nonumber
\end{eqnarray}
This can be rewritten as,
\begin{eqnarray}\label{genfn}
  \Phi &=&1+ \delta\sum_{n\neq0}
\left[(1+\delta^2)\sinh^2{n\mu_0}\right. \\
&&\quad \left.+\delta\cos\theta\sinh{2n\mu_0}
        +\delta^2 \right]^{-1/2}\nonumber
\end{eqnarray}
where $\delta\equiv a/R=M/4R\Sigma_1$, and where we have eliminated $a$
by using (\ref{Mass}).

\begin{figure}
\epsfxsize=200pt \epsfbox{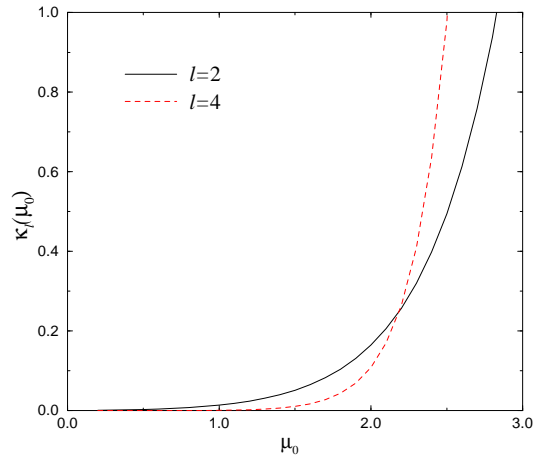}
\caption{The $\kappa_\ell$ coefficient as a function of $\mu_0$.}
\label{fig10}
\end{figure}

The sum in (\ref{genfn}) is recognized as the generating function for
the Legendre polynomials, so  the conformal factor can be rewritten as
\begin{equation}
   \Phi = 1+2\sum_{\ell=0,2,4...} \kappa_\ell(\mu_0)
   \left(M/R\right)^{\ell+1}P_\ell(\cos\theta) \quad .
\end{equation}
The only $\mu_0$ dependence occurs in the $\kappa_\ell$ coefficients,
\begin{equation}\label{kapdef}
       \kappa_\ell(\mu_0)
\equiv \frac{1}{\left(4\Sigma_1\right)^{\ell+1}} \sum_{n=1}^\infty
        \frac{(\coth{n\mu_0})^\ell}{\sinh{n\mu_0}}
\end{equation}
and this dependence is shown in figure \ref{fig10}.

If the $\ell=0$ term in the sum for $\Phi$ is explicitly evaluated
(note that $\kappa_0(\mu_0)=1/4$) the result is
\begin{equation}
    \Phi = \left(1+\frac{M}{2R}\right){\cal F},
\end{equation}
with
\begin{equation}\label{Fdef}
       {\cal F}
\equiv 1+2 \left( 1+\frac{M}{2R}\right)^{-1}\!\!
       \sum_{\ell=2,4...} \!\!\kappa_\ell(\mu_0)
       \left(M/R\right)^{\ell+1}P_\ell(\cos\theta).
\end{equation}

Since $R$ is analogous to the radial coordinate in isotropic
coordinates for a spherical geometry, we introduce a
Schwarzschild-like radial coordinate $r$ by the transformation
\begin{equation}\label{rtrans}
   R = \frac{1}{4}\left(\sqrt{r}+\sqrt{r-2M}\right)^2
\end{equation}
that relates the two coordinates in the Schwarzschild geometry.
With this transformation we arrive, finally, at
\begin{equation}\label{M3}
  ds^2_{\rm Misner} = {\cal F}(r,\theta)^4
  \left(\frac{dr^2}{1-2M/r}+r^2d\Omega^2\right)
\end{equation}
in which the $r$ and $\theta$ dependence of ${\cal F}$
are given by (\ref{Fdef}) and (\ref{rtrans}).
It is easy to show that a choice of time coordinate $t$
can be made so that the 4-geometry generated by the Misner
initial data takes the form
\begin{eqnarray}\label{SchForm}
  ds^{2} =&& -\left(1-\frac{2M}{r}\right)dt^{2}\\
&&+{\cal F}(r,\theta)^4
  \left[ \frac{dr^{2}}{1-2M/r}+r^{2}d\Omega^{2} \right]\nonumber
\end{eqnarray}
at $t=0$ with all first time derivatives of the metric vanishing.

The function ${\cal F}$ may therefore be thought of as containing the
mathematical description of how the Misner geometry initially deviates
{}from the Schwarzschild geometry.

In treating (\ref{SchForm}) as a distorted Schwarzschild geometry
we are taking the range of $r$ to be $2M<r<\infty$ and hence, by
(\ref{rtrans}), our range of coordinates must include all points
with $R>M/2$.  But according to the transformation
in (\ref{transforms}) the values of $R$ at $\eta=0$
are limited (by the condition $|\mu|<\mu_0$) to values for which
\begin{equation}\label{Rrange}
  R\geq a\coth{\mu_0/2} = (M/4\Sigma_1)\coth{\mu_0/2}\quad .
\end{equation}
Thus our coordinates can reach down to $R=M/2$, while corresponding to
$|\mu|\geq\mu_0$ only if $\coth{\mu_0/2}\leq2\Sigma_1$, and this
condition turns out to be satisfied only for $\mu_0\leq1.191$.
We ignore this restriction and find (see below) that the
perturbation approach gives reasonably accurate answers for values of
$\mu_0$ somewhat larger than 1.191. The explanation is that all the
strong non-spherical deviations are very close to the horizon and end
up not having important consequences for emitted radiation.

Note that the Misner geometry with the $\kappa_\ell$ set to zero would
be precisely a $t=$ const. slice of the Schwarzschild geometry. The
deviations from Schwarzschild are determined by the $\kappa_\ell$'s in
(\ref{Fdef}). If they are sufficiently small then the Misner geometry
may be considered initially to be a perturbed Schwarzschild
geometry. Since the Schwarzschild geometry is stable, sufficiently
small initial perturbations will remain perturbations, and the
spacetime generated from the Misner initial data will be a perturbed
Schwarzschild spacetime. A crucial point is that the form of the
initial data interior to the initial horizon cannot affect the
evolution of the exterior. We can, therefore, view the generation of
outgoing waves --- a process confined to the exterior of the horizon
--- as nearly spherically symmetric if the initial data is nearly
spherically symmetric only outside the horizon.  More specifically: if
$M/R$ is of order unity, or smaller, outside the initial horizon, and
if the $\kappa_\ell(\mu_0)$ coefficients are small, then the evolved
spacetime outside the horizon will be a perturbed Schwarzschild
geometry.

This picture makes sense only if the coefficients ${\kappa}_{\ell}$'s
become small when $\mu_0\rightarrow 0$, and if this limit corresponds
to the ``close limit'' in which the separation of the two black holes
vanishes.  To show the former  we notice that
$(\coth{n\mu_0})^{\ell}/\sinh{n\mu_{0}}\approx(n\mu_0)^{\ell+1}$ for
$n\mu_0<<1$. We assume that this approximation is valid for small
$\mu_{0}$, and we keep terms in summations only up to
$N\sim1/\mu_{0}$. With these approximations we have, for $\ell\geq1$.

\begin{equation}
       \sum_{n=1}^\infty \frac{(\coth{n\mu_0})^\ell}{\sinh{n\mu_0}}
\approx \frac{1}{({\mu_0}^{\ell+1})}\sum_1^N\frac{1}{n^{\ell+1}}
      \approx \frac{\zeta(\ell+1)}{{\mu_0}^{\ell+1}}
\end{equation}
and similarly
\begin{equation}
       \Sigma_1\equiv\sum_{1}^\infty\frac{1}{\sinh{n\mu_0}}
\approx \frac{1}{\mu_0}\sum_1^N\frac{1}{n}
\approx \frac{\ln{N}}{\mu_0}
      \approx\frac{|\ln{\mu_0}|}{\mu_0}\quad .
\end{equation}
These approximations in (\ref{kapdef}) give us
\begin{equation}
  \kappa_\ell(\mu_0)
  \approx \frac{\zeta(\ell+1)}{|4\ln{\mu_0}|^{\ell+1}}\quad .
\end{equation}
Although the steps used to derive this result were very rough
approximations, the result is in good agreement with numerical
values of the $\kappa_\ell(\mu_0)$\,s.

We turn now to the question of the separation of the throats in the
limit of small $\mu_0$. The proper distance $L$ between the throats
(more specifically, the distance from $\mu=-\mu_0$ to $+\mu_0$
along the line $\eta=\pi$) can be written as the sum\cite{Lindquist63}
\begin{equation}
 L = 2a\left(1+2\mu_0\sum_1^\infty \frac{n}{\sinh{n\mu_0}} \right)\quad .
\end{equation}
Here it will be more useful to use an equivalent result in which
$L(\mu_0)$ is given implicitly in terms of $K$ and $E$, the complete
elliptic integrals of the first and second kind:
\begin{equation}
  L = (4a/\pi)K \left[(1-k^{2})^{1/2}\right] E(k)
\end{equation}
\begin{equation}
 \mu_{0} = \pi K(k)/K\left[(1-k^{2})^{1/2}\right]\quad .
\end{equation}
If $\mu_0$ is small, then this forces $k\ll 1$ so that
$E(k)\approx K(k)\approx\pi/2$, and
\begin{equation}
 L = (\pi^{2}a/\mu_{0}) \left[1+{\cal O}(k^{2})\right]
\end{equation}
with $k\approx4\exp{(-\pi^{2}/2\mu_{0})}$. We then have
\begin{equation}
        \frac{L}{M}
\approx \frac{\pi^{2}}{4\mu_{0}\Sigma_1}
\approx \frac{\pi^2}{4|\ln{\mu_0}|}\quad ,
\end{equation}
and hence the separation decreases as $\mu_0$ decreases.

We can then consider $\epsilon\equiv1/|\ln{\mu_0}|$ to be a
perturbation parameter, with which we can evolve the Misner
initial using perturbation theory. The form of the series
in (\ref{Fdef}) appears to give a multipole decomposition
of the deviations from spherical symmetry.  This is not quite
true.  The metric perturbations involve ${\cal F}$ raised
to the fourth power. When this is done terms of different
multipolarity in ${\cal F}$ mix to give each multipole of
${\cal F}^4$. The quadrupole term in ${\cal F}^4$, for example,
will contain contributions from the product of the
$\ell=2$ and $\ell=4$ terms in ${\cal F}$, from the square of
the $\ell=2$ term, etc. The total quadrupole term will contain
products of the form $\kappa_2\kappa_4,
\kappa_2^2,\kappa_8\kappa_2\kappa_4$, and so forth.
This complexity causes no difficulty in practice.
Of all the contributions to the quadrupole in ${\cal F}^4$,
that of lowest order is the {\cal O}($\epsilon^3$) term linear
in $\kappa_2$; the next lowest order contribution
${\cal O}(\epsilon^6$) is that from $\kappa_2^2$.
If we are evolving the initial data with the evolution
equations of linear perturbation theory for the quadrupole
then we must keep {\it only} the part of ${\cal F}^4$ that
is linear in $\kappa_2$. It would be inconsistent to keep
any higher order terms since we are ignoring the contributions
of order ${\cal O}(\epsilon^6)$ due to the nonlinear evolution
of the initial data. A variation of the same argument applies
in the case of other multipoles. If we are using the $\ell$-pole
linear equations to evolve the $\ell$-pole initial data,
we must keep only the term in ${\cal F}^4$ that is linear in
$\kappa_\ell$. All other $\ell$-pole contributions will be higher
order in $\epsilon$. Thus if we are only interested, for each
$\ell$-pole order, in the terms that can be treated with linearized
perturbation theory, we may write ${\cal F}^4$ as
\begin{equation}\label{Ys}
           {\cal F}^4
\approx  1+8\,(1+\frac{M}{2R})^{-1}\!\!\sum_{\ell=2,4...}
           \kappa_\ell(\mu_0) \left(M/R\right)^{\ell+1}
           P_\ell(\cos\theta),
\end{equation}
where
\begin{equation}
P_\ell(\cos\theta)=
           \sqrt{\frac{4\pi}{2\ell+1}}Y_{\ell0}(\theta,\phi) \quad .
\end{equation}

To evolve these even parity perturbations we use the formalism and the
notation of Moncrief~\cite{Moncrief74}.  From Eqs.~(5.1)--(5.8)
of that paper, applied to the perturbed spacetime of our
Eqs.~(\ref{SchForm})--(\ref{Ys}), we find that Moncrief's perturbation
functions $h_1$ and $G$ vanish, and
\begin{equation}\label{H2K}
  H_2=K = {\cal G}_\ell(r; \mu_0)= 8\sqrt{\frac{4\pi}{2\ell+1}}
          \kappa_\ell(\mu_0)\frac{(M/R)^{\ell+1}}{1+M/2R}
\end{equation}
and Moncrief's $q_1$ is
\begin{eqnarray}
  q_1 =&& 2r \left(1-\frac{2M}{r}\right)
  \left[ {\cal G}_\ell - \sqrt{1- \frac{2M}{r}} \frac{d}{dr}
  \left( \frac{r{\cal G}_\ell}{\sqrt{1-\frac{2M}{r}}} \right)
  \right] \nonumber\\
&&+ \ell(\ell+1)r{\cal G}_\ell \quad .
\end{eqnarray}
We then define, for any $\ell$
\begin{equation}\label{psidef}
       \psi_{\rm pert}
\equiv \frac{(\ell+2)(\ell-1)}{(\ell+2)(\ell-1)+6M/r}q_1\quad .
\end{equation}
For $\ell=2$ note that $\psi_{\rm pert}$ is identical
to the $\psi$ of \cite{Price94a}.
Note also that the definitions here for the metric perturbations are
closely related to those in Eqs.(\ref{k1eq})-(\ref{psi eq}), except that
the normalization used for $\psi$ is different (see below).

At this point it should be observed that in Eqs.~(\ref{H2K}) to (\ref{psidef})
the metric perturbations are proportional to $\kappa_\ell(\mu_0)$ and
that there is no other $\mu_0$ dependence in the perturbations. We may
therefore view these expressions, for each $\ell$-pole moment, as
first-order perturbation theory in $\kappa_\ell(\mu_0)$. From a formal
point of view we may consider that we have a family of spacetimes
parameterized by $\mu_0$, and we are approximating the metric by
\begin{equation}\label{fstorder}
g_{\alpha\beta}\approx \left.g_{\alpha\beta}\right|_{\kappa_\ell(\mu_0)=0}+
\left.\frac{\partial
g_{\alpha\beta}}{\partial \kappa_\ell(\mu_0)}
\right|_{\kappa_\ell(\mu_0)=0}\times\kappa_\ell(\mu_0)\ .
\end{equation}
It is important to realize that we could just as well do perturbation
theory using another expansion parameter $\tilde{\kappa}_\ell(\mu_0)$
that agrees to linear order with $\kappa_\ell(\mu_0)$; we might for
example use $\tilde{\kappa}_\ell(\mu_0) =
\kappa_\ell(\mu_0)+\left[\kappa_\ell(\mu_0\right]^2$. In terms of this
parameter our first-order approximation would be
\begin{displaymath}
g_{\alpha\beta}
\approx \left.g_{\alpha\beta}\right|_{\tilde{\kappa}_\ell(\mu_0)=0}+
\left.\frac{\partial g_{\alpha\beta}}
{\partial \tilde{\kappa}_\ell(\mu_0)}
\right|_{\tilde{\kappa}_\ell(\mu_0)=0}\times\tilde{\kappa}_\ell(\mu_0)
\end{displaymath}
\begin{equation}\label{tilde}
=\left.g_{\alpha\beta}\right|_{\kappa_\ell(\mu_0)=0}+
\left.\frac{\partial
g_{\alpha\beta}}{\partial \kappa_\ell(\mu_0)}
\right|_{\kappa_\ell(\mu_0)=0}\times\tilde{\kappa}_\ell(\mu_0)
\end{equation}
The perturbations here differ from those  in (\ref{fstorder}) by the factor
$\tilde{\kappa}_\ell(\mu_0)/\kappa_\ell(\mu_0)$.  The approximation
given by first-order perturbation theory, then, is dependent on our
choice of expansion parameter.  This, of course, is a reflection of
the fact that first-order perturbation theory must be uncertain to
second-order in the expansion factor, but it should be kept in mind
that the remarkably good predictions of perturbation theory at
unexpectedly large values of $\mu_0$ are to some extent due to the
particular (though rather natural) use of $\kappa_\ell(\mu_0)$ as the
effective expansion parameter.

The function $\psi_{\rm pert}$ is evolved with the Zerilli equation
\begin{equation}
   \frac{\partial^2\psi_{\rm pert}}{\partial t^2}
 - \frac{\partial^2\psi_{\rm pert}}{\partial r^{\ast 2}}
 - V(r^\ast)\psi_{\rm pert}=0
\end{equation}
in which $r^\ast $ is a ``tortoise'' coordinate such that
$r^\ast \rightarrow\infty$ corresponds to spatial infinity
and $r^\ast \rightarrow -\infty$ corresponds to the horizon,
\begin{equation}
  r^\ast \equiv r+2M\ln{\left(\frac{r}{2M}-1\right)}\quad.
\end{equation}
and the Zerilli potential is given by,
\begin{eqnarray}
  V(r^\ast )=&& \left(1-\frac{2M}{r}\right)
  \left\{\frac{1}{\lambda^2}
  \left[\frac{9M^3}{2r^5}-\frac{3M}{r^3}
  \left(1-\frac{3M}{r} \right)\right]\right.\nonumber\\
 &&\left.+ \frac{6}{r^2\lambda}\right\}
\end{eqnarray}
and $\lambda\equiv1+3M/2r$.

\begin{figure}
\epsfxsize=200pt \epsfbox{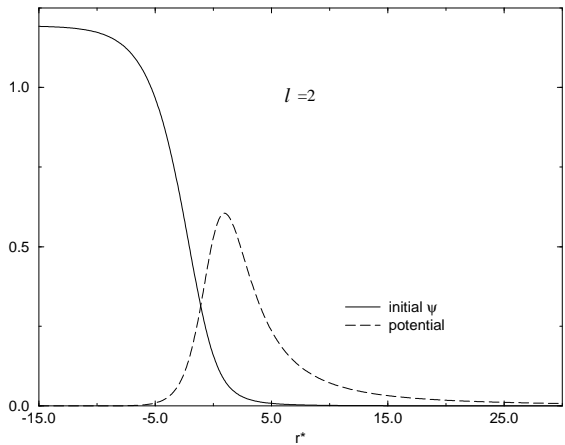}
\epsfxsize=200pt \epsfbox{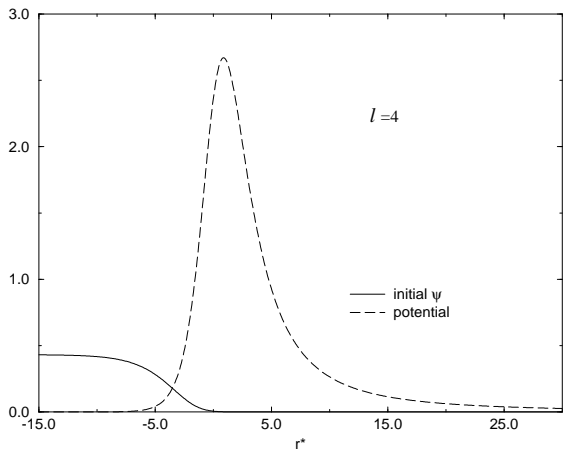}
\caption{We show the initial perturbation waveform, computed
at $t=0$ directly from the Misner initial data, and the Zerilli
potential for $\ell=2,4$.}
\label{X}
\end{figure}

The initial form of $\psi_{\rm pert}$, given by
Eqs.~(\ref{H2K})-(\ref{psidef}), is shown in Fig.~\ref{X} along with the
potential for the Zerilli equation. Both the and $\ell=2$ and $\ell=4$
figures are shown.

It is straightforward to show that, in terms of $\psi_{\rm pert}$,
the radiation power is given by
\begin{eqnarray}\label{powerdistr}
   \frac{d{\rm Power}}{d\Omega}
 =&& \frac{1}{16\pi}\left(\frac{d\psi_{\rm pert}}{dt}\right)^2
   \left[\frac{(\ell-2)!}{(\ell+2)!}\right]^2\nonumber\\
&&\times
   \left(\frac{\partial^2Y_{\ell0}}{\partial\theta^2}
   -\cot\theta\frac{\partial Y_{\ell0}}{\partial\theta}\right)^2
\end{eqnarray}
and therefore,
\begin{equation}\label{power}
  {\rm Power} = \frac{1}{16\pi}
  \left(\frac{d\psi_{\rm pert}}{dt}\right)^2
  \frac{(\ell-2)!}{(\ell+2)!}
\end{equation}
Since the waveforms and power depend on $\mu_0$ only through
factors of $\kappa_\ell(\mu_0)$,
we can compute with $\kappa_\ell$ set to unity,
and get the correct results for any $\mu_0$ by multiplying waveforms
and power expressions respectively by $\kappa_\ell(\mu_0)$ and
$\kappa_\ell(\mu_0)^2$.

The above formalism has been used to generate the waveforms and the
energies that are presented below and compared to the results of numerical
relativity. In the comparisons we must take
account of the fact that different normalization conventions than
those above have been used to define the wave function
$\psi_{\rm num}$. The comparisons  will be made
by converting the perturbation waveform according to
\begin{equation}
  \psi_{\rm num}
= \sqrt{2\frac{(\ell-2)!}{(\ell+2)!}}\psi_{\rm pert}
\end{equation}

\subsection{Comparison of results}
\label{sec:results}

\begin{figure}
\epsfxsize=200pt \epsfbox{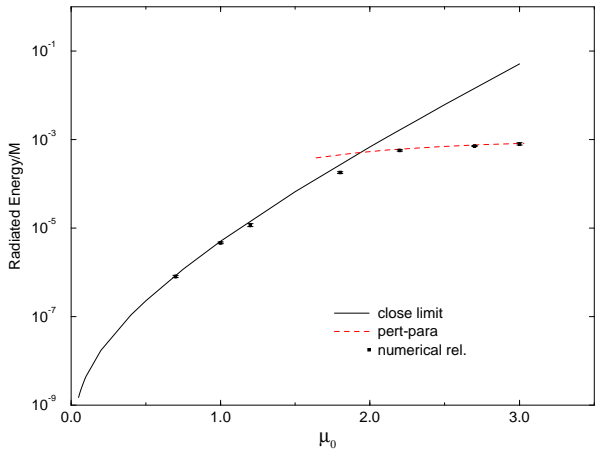}
\caption{Comparison of the close approximation energy with the full
numerical results and the particle-membrane approximation}
\label{fig:VI-1}
\end{figure}

Probably the most important comparison to be made is the radiated
energy computed by the different methods for dealing with the
outgoing radiation.  In Fig.~\ref{fig:VI-1} this comparison is given.
The numerical relativity supercomputer results are shown along with
error bars indicating the range of energies found by extracting
waveforms at different radii. There are no analogous formal errors
for the close-limit or the perturbation-paradigm approach.

The results show remarkable agreement (already noted
in \cite{Price94a}) between the close-limit prediction and
the results of fully nonlinear numerical relativity.
This agreement is reasonably good even up to $\mu_0=2$,
where the energy results differ by only 15\%. Above $\mu_0=2$
the results quickly diverge, with the close-limit results seriously
overestimating the radiated energy. For these cases, however,
the perturbation-paradigm method gives excellent agreement
with computed results (as has already been noted
in \cite{Anninos93b,Anninos94b}).

\begin{figure}
\epsfxsize=200pt \epsfbox{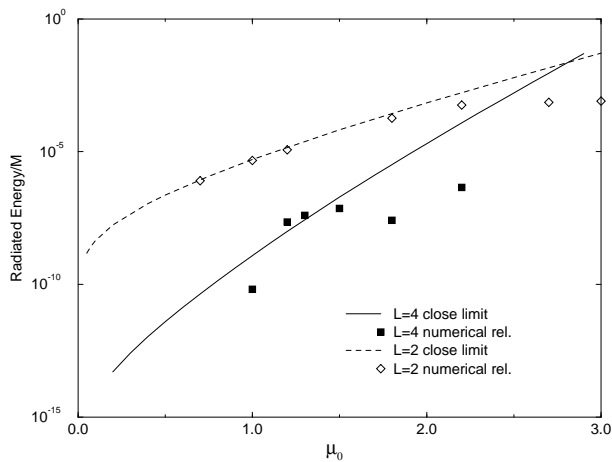}
\caption{Comparison of energies for both the $\ell=2$ and  $\ell=4$ modes.
The numerical relativity results are denoted by symbols showing the specific
case computed, while the close approximation results are denoted by lines.}
\label{fig:VI-2}
\end{figure}

The energies shown in Fig. \ref{fig:VI-1} are strongly dominated by the
quadrupole contribution.  It is instructive to look at the
$\ell=4$ energy predictions of numerical relativity and of the
close limit.  In the case of the close limit, the $\ell=4$
energy is computed in the manner explained in Sec.~IVA.  In the
numerical relativity results the $\ell=4$ results must be
extracted by fitting the angular pattern as explained in Sec.~II.
The numerical $\ell =4$ results lack the clear trend,
seen in Fig.~\ref{fig:VI-2}, of agreement at small $\mu_0$, monotonically
growing worse with increasing $\mu_0$.

\begin{figure}
\vspace{-2cm}
\epsfxsize=200pt \epsfbox{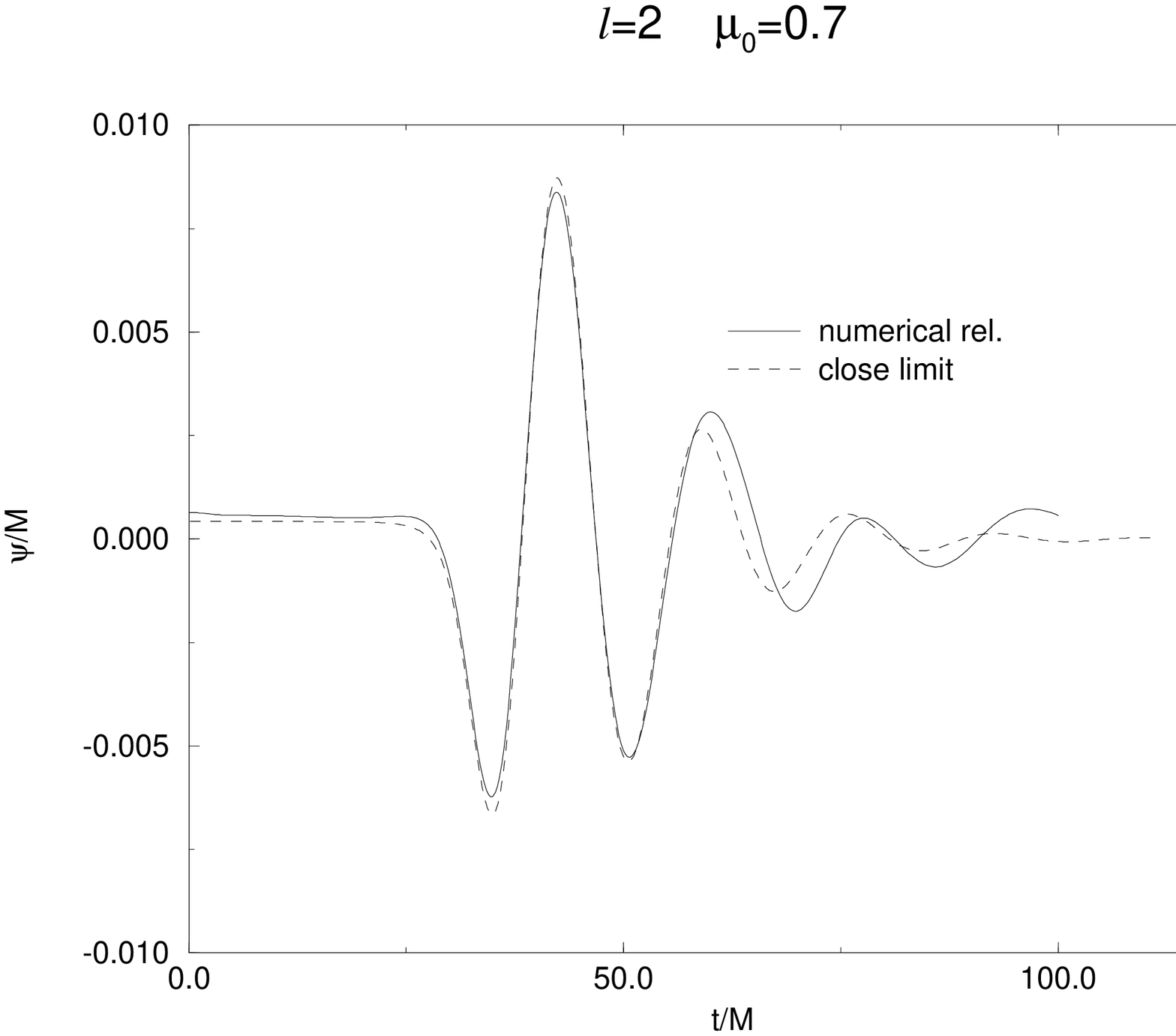}
\epsfxsize=200pt \epsfbox{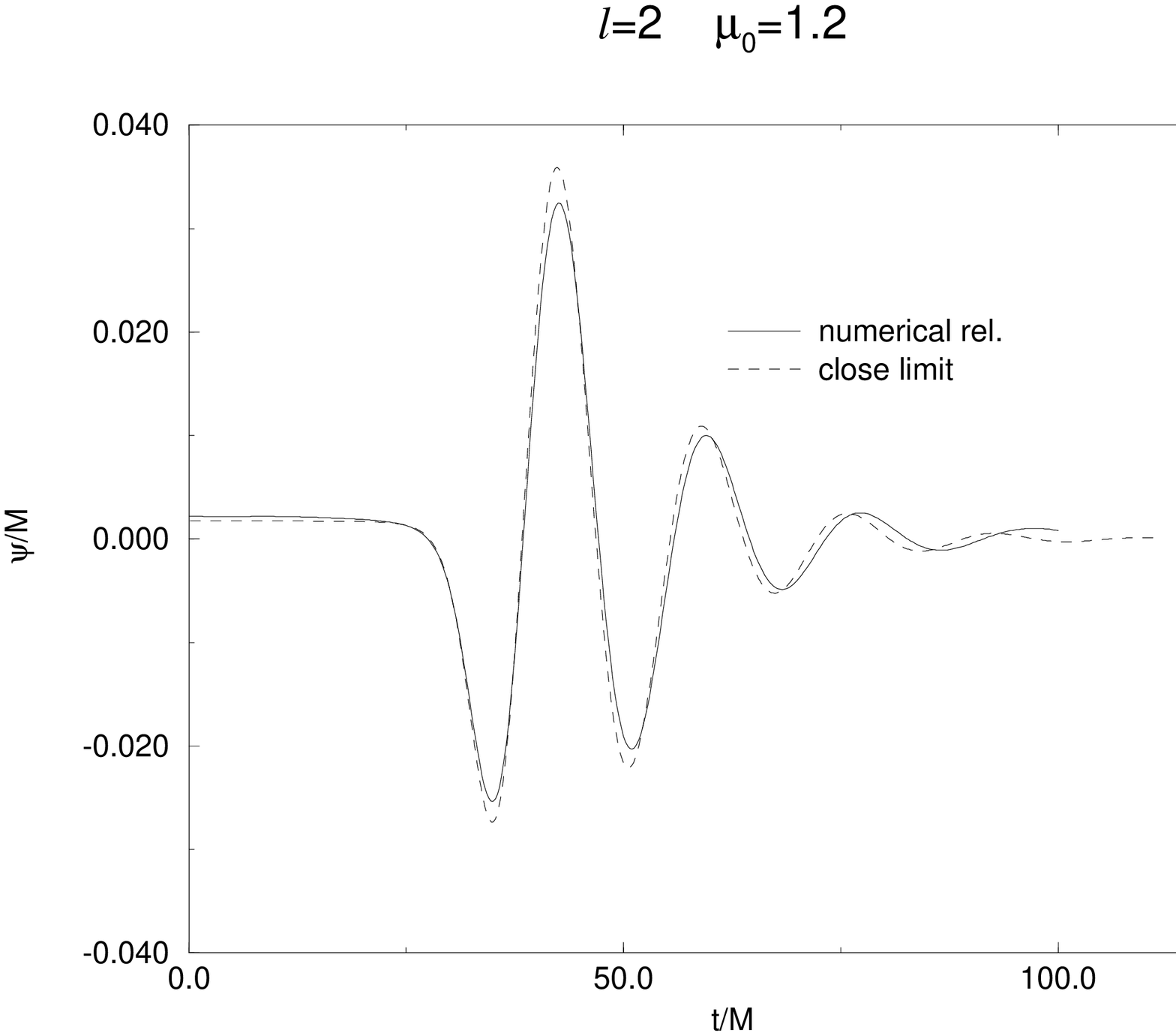}
\epsfxsize=200pt \epsfbox{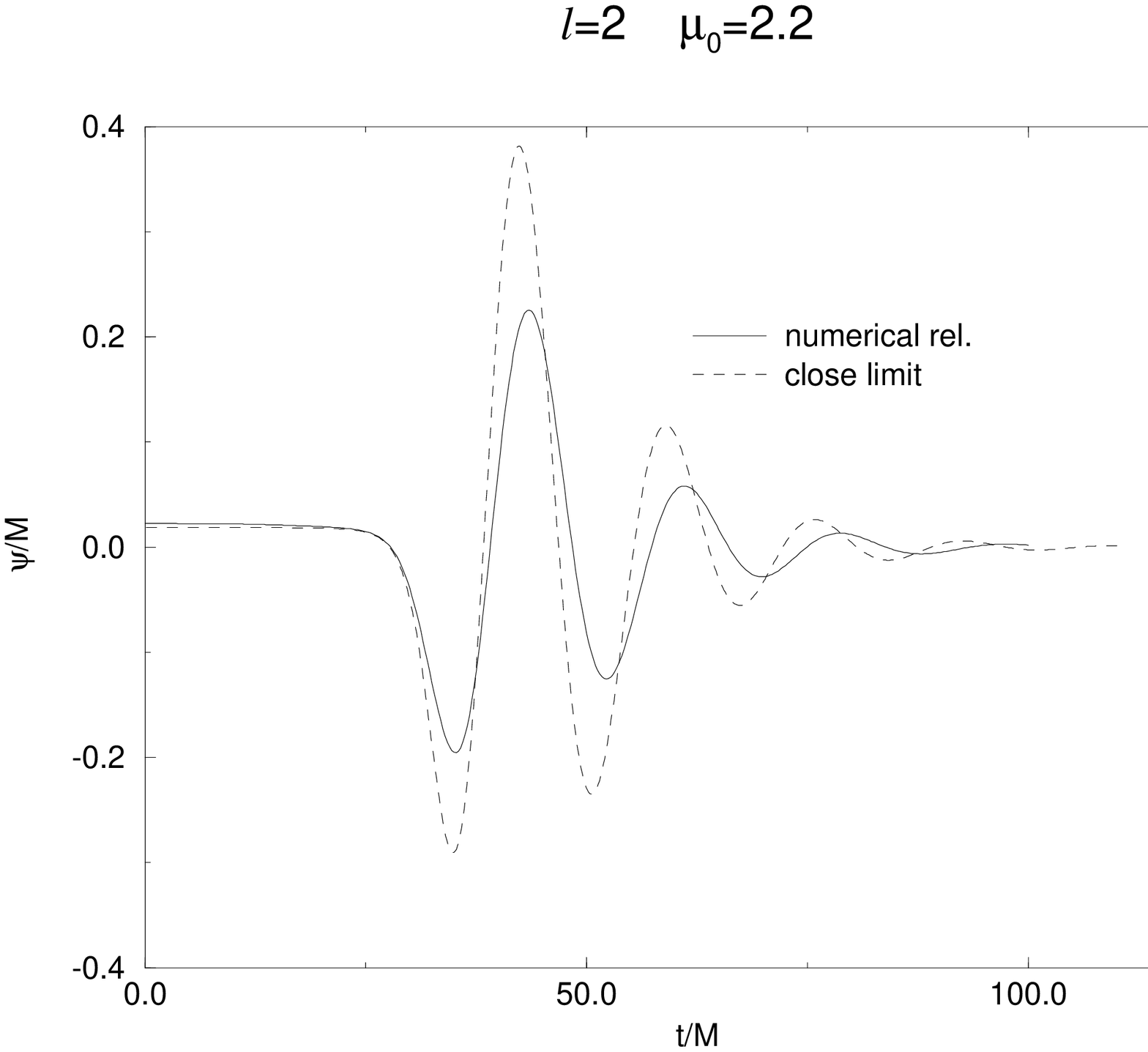}
\caption{Comparison of the $\ell=2$ waveforms obtained by the
close approximation and those obtained by full numerical relativity for
several cases.}
\label{fig:VI-3}
\end{figure}

A better understanding of the sources of disagreement of the method
follows from an examination of the ``waveforms,'' the time dependence
of the perturbation function $\psi$ at a constant radius.  Figure
\ref{fig:VI-3}  shows a series of waveforms ``observed'' at
$r=25M$ (where here and below $M$ refers to the ADM mass of the
initial data), and show the astonishingly good agreement at small
$\mu_0$, agreeing in form not only in the region dominated by
``ringing'' at the quasinormal frequency, but also agreeing at early
times.  As $\mu_0$ increases, the curves continue to agree in general
character, but disagree in the amplitude of the quasinormal ringing.
Only a single interesting feature (aside from the agreement!) appears
in these curves.  There is a consistent phase drift between the
close-limit waveforms, and the numerical relativity waveforms; the
late-time quasinormal ringing of the numerical relativity waveforms
has a frequency too low by 10--20\%.  (The correct values of the
quasinormal frequencies are well known from other calculations. See
e.g.,~\cite{Leaver85}).  As pointed out in Sec.~II, this suggests that
there is a systematic effect in the present numerical relativity
computations causing this phase drift and leading to underestimates of
radiated energy.  Any increase in the numerical computed energy at
large $\mu_0$ will, of course, improve the agreement with the
close-limit estimates.

\begin{figure}
\vspace{-2cm}
\epsfxsize=200pt \epsfbox{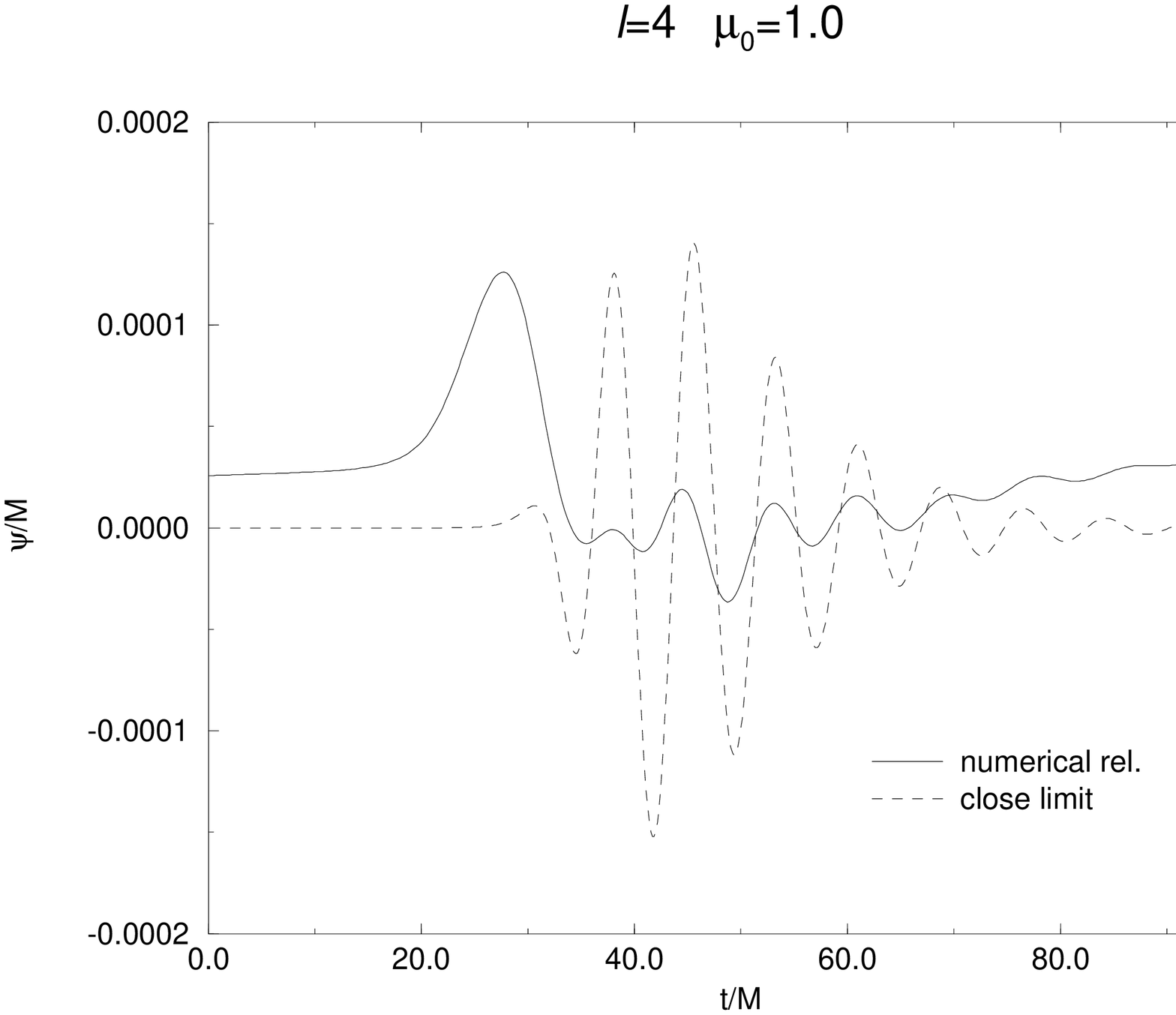}
\epsfxsize=200pt \epsfbox{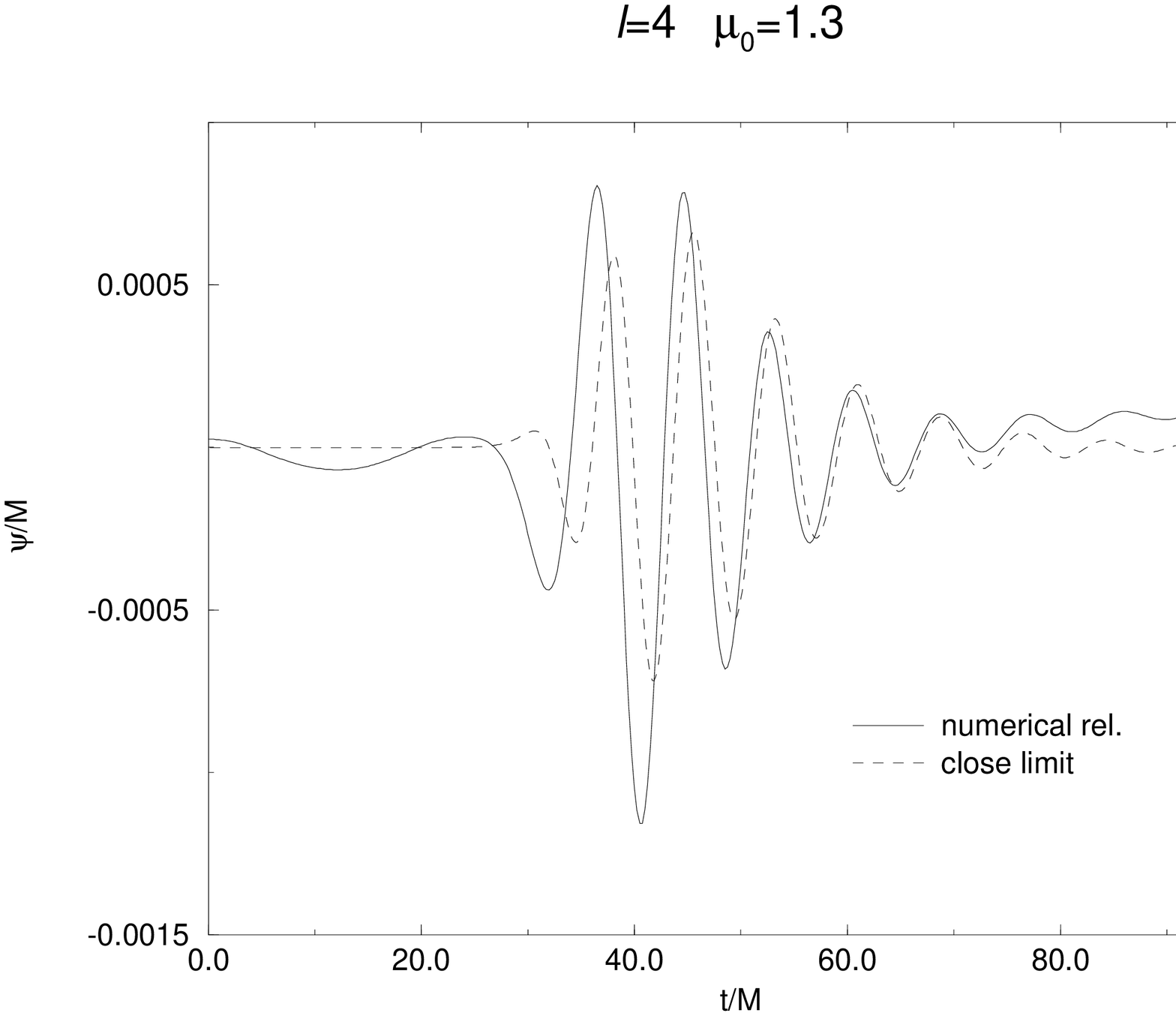}
\epsfxsize=200pt \epsfbox{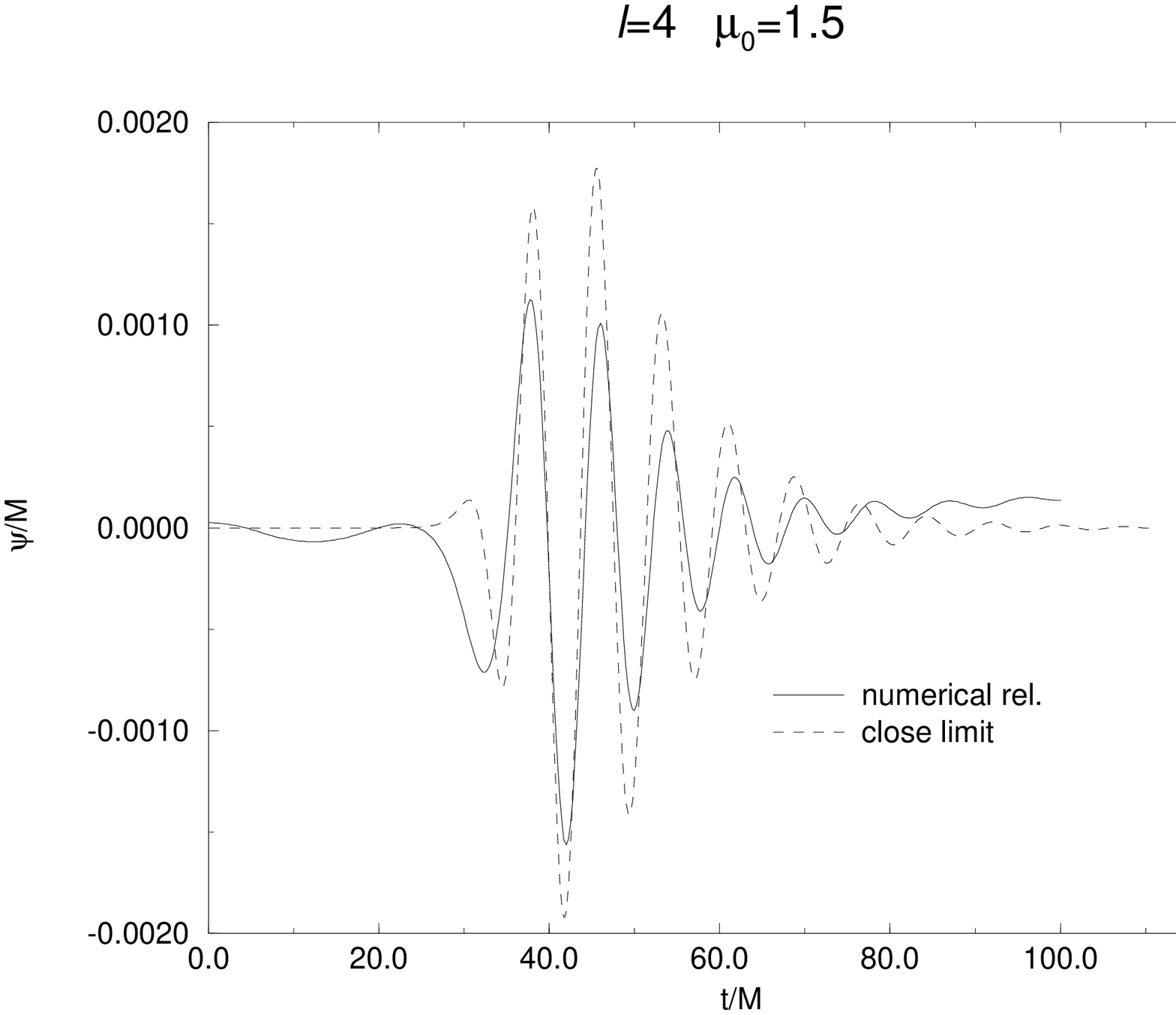}
\caption{Comparison of the waveforms for $\ell=4$
{}from the close approximation and from numerical relativity.}
\label{fig:VI-4}
\end{figure}

The comparison of $\ell=4$ waveforms, at $r=25M$, shown in Fig.~\ref{fig:VI-4},
clarifies the trends in Fig. \ref{fig:VI-2}.  For small $\mu_0$
numerical errors in the multipole extraction scheme result in
waveforms which are clearly in error.  Not only do the waveforms lack
the expected quasinormal ringing, they have a non-physical trend at
late times.  These errors are so large simply because at small $\mu_0$
the radiation is overwhelmingly dominated by the quadrupole part.  At
$\mu_0 =1.0$, for example, the $\ell=4$ contribution is only 1\% of
the amplitude of the wave.  As pointed out in Sec.~II, extraction of
this very small part is very sensitive to numerical noise.  As
$\mu_0$ increases the relative size of the $\ell=4$ contribution
increases, and the numerical error involved in extracting it
decreases.  For $\mu_0\approx 1.2-1.3$ the numerical errors are small
enough so that the waveforms show overall reasonable agreement (along
with curious features at early and late times).  At larger values of
$\mu_0$ it would be expected that the extracted waveforms would
continue to be more accurate, but the lack of a monotonic increase in
energy shows that the errors in the waveforms are still very large.
Though the large errors make conclusions uncertain, the results
suggest that the linearized approach in the close-limit method has a
smaller range of validity for $\ell=4$ than for $\ell=2$.  Despite
this somewhat smaller range of validity the point should not be missed
that at the present state of the art in numerical relativity, the
close-limit waveforms and energies, for a range of $\mu_0$, are the
only reliable estimates available for $\ell > 2$.

\begin{figure}
\epsfxsize=200pt \epsfbox{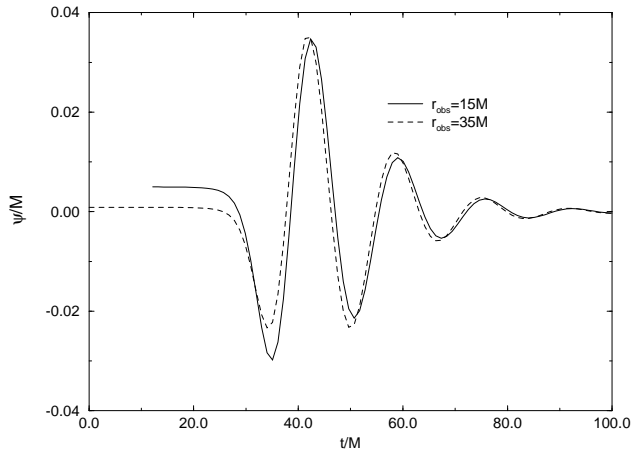}
\epsfxsize=200pt \epsfbox{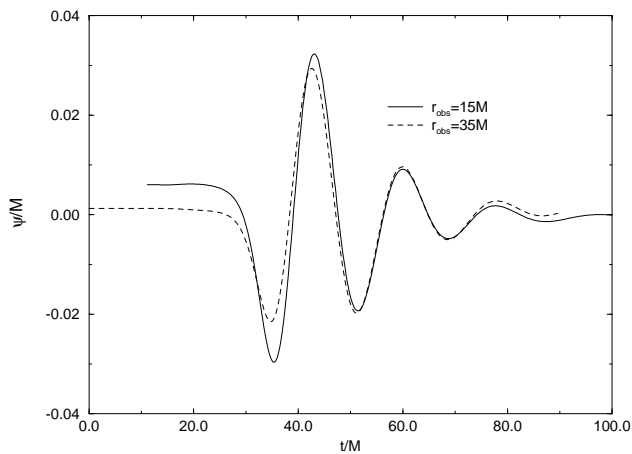}
\caption{(a,b)Comparison of the waveforms ``observed'' at different radii for
the close approximation and full numerical cases for $\mu_0=1.2$.  (a)
shows the results of the close approximation and (b) shows the full
numerical relativity results.  The phases have been adjusted for comparison
of the waveform shape at the different radii.}
\label{fig:VI-5}
\end{figure}

The waveforms presented above have been ``observed'' at radius
$r=25\,M$. That is, the figures showed $\psi(r=25\,M, t)$. To explore
the sensitivity of the comparisons to different observation radii, in
Figs.~\ref{fig:VI-5} we present $\mu_0=1.2$ waveforms observed at
different radii.  Close-limit waveforms are shown in
Fig.~\ref{fig:VI-5}a, and those of numerical relativity in
Fig.~\ref{fig:VI-5}b.

In these figures the phases of all waveforms have been made compatible
with each other, and with Figs.~\ref{fig:VI-3} and \ref{fig:VI-4}.
Each waveform has been shifted by a time equal to the difference
in observation values of $r^\ast$, and the value of $r^\ast$ equivalent
to $r=25\,M$. In Figs.~\ref{fig:VI-5} we see that these phase
corrected waveforms agree very well at late times, but have somewhat
different initial shapes. At early times the waveform observed at
smaller radius is larger than if observed at larger radius.
This, of course, is a manifestation of the fact that at very early
times what we are seeing is essentially the initial data,
which -- unlike the outgoing radiation -- falls off in radius.
This effect, the presence of a non-radiative part of the waveform,
is significant up to around the first peak of quasinormal ringing.
It is important to note that the differences in waveforms observed
at different radii are present in both the close-limit and
the numerical relativity results. In fact, the differences
in the waveforms for different radii are much larger than
the differences between the waveforms computed by numerical
relativity and by the close-limit approximation.

\begin{figure}
\epsfxsize=200pt \epsfbox{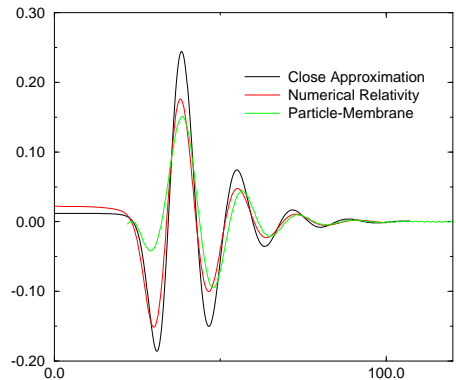}
\caption{Comparison of the waveforms for the
particle-membrane approximation, numerical relativity and the close
approximation for the case $\mu_0=2.0$. These three calculations,
containing no adjustable parameters, agree remarkably well, although
for this separation the close approximation is near the limit of its
applicability. }
\label{fig:VI-6}
\end{figure}

The waveform comparison to this point has been between the
computations of numerical relativity and the close-limit
approximation. We now tie this work together with the particle-membrane
calculation described in section~\ref{sec:timesym}.  Although the two
approximations are based on different limits, as we have seen
they nearly overlap if the two holes are not too far or too close.
In Fig.~\ref{fig:VI-6} we show results from all three approaches
we have used in this paper for the case $\mu_0=2.0$. The solid line
shows the close approximation, the dashed line shows the results
{}from the full numerical relativity calculation, and the dotted line
shows the particle-membrane calculation for the time symmetric
particle perturbation.  It is important to emphasize that
the three graphs shown contain no adjustable parameters.

The waveforms show good agreement at late retarded time, when the
behavior is dominated by quasinormal ringing.  The waveform computed
{}from the close limit has a higher amplitude, as this approximation is
not as accurate for holes at such a large separation parameter.  There
is, however, a more significant difference in the shape of the early
waveforms.  Here it is the particle-membrane approach that differs
{}from the other calculations.
As discussed in
Sec.~\ref{sec:timesym}, for this $\mu_0=2.0$ case,
the extraction of a waveform is difficult and is sensitive to the
details of the extraction procedure.
Overall, in view of  the very different nature of the three
calculations, and their different regimes of validity, the agreement
among these results is remarkable.

\section{Discussion and Conclusion}
\label{sec:discussion}

Conventional wisdom holds that events involving colliding black
holes are so nonlinear that they can only be studied using numerical
relativity.  We have shown in this paper that this is not necessarily
the case.  In particular, we have shown that the head-on collision
of two black holes (whose starting point is the Misner initial data)
is amenable to semi-analytic treatments giving reliable results in a
considerable range of the parameter space.

In the Misner data, the initial separation of the two black holes is
determined by the value of the parameter $\mu_0$. At present, our
numerical techniques are capable of evolving the Misner data, and of
extracting waveforms, for the range $0.7 \leq \mu_0 \leq 3.0$,
corresponding to initial proper separations $L$ ranging from $1.51 M$
to $7.92 M$. For $\mu_0<1.8$ ($L < 3.38 M$, corresponding to $r_0 =
2.82 M$), the black holes are
surrounded by a common event horizon.  In such cases the perturbative
treatment (``close limit'') described in Sec.~IV provides a reliable
framework for approximating the true evolution of the spacetime, and
the resulting waveforms match nicely those obtained numerically.  For
$\mu_0 > 1.8$, the black holes are truly distinct, and we find that
the waveforms can be well reproduced using the ``particle-membrane''
treatment described in Sec.~III.  This approach can also be used to
``predict'' the waveforms produced during large-$\mu_0$ collisions,
cases which should be amenable to numerical calculation in the near
future.

We note that in Sec.~III B and IV B, the comparisons between the
numerical and semi-analytic waveforms are made without using a single
adjustable parameter. For the $\ell=2$ component of the waveforms
(the dominant component), the agreement between the numerical and
semi-analytic approaches is remarkable. For $\ell=4$, extraction
of the waveforms from the numerical data is less accurate,
especially for smaller values of $\mu_0$. As a result, and
as shown in Sec.~IV B, the numerical and semi-analytic $\ell=4$
waveforms differ significantly when $\mu_0 \leq 1.0$. For
larger values of $\mu_0$, the $\ell=4$ waveforms agree well
in wavelength and in phase, but less so in amplitude. This is
illustrated in Figs.~\ref{fig:VI-4}. To explain this, we
note that the waveform amplitude is more sensitive than other
waveform attributes (such as wavelength) to the choice of
parameters (such as resolution) used in the numerical
calculation.

What have we learned about the Misner data through the semi-analytic
studies?  For gravitational waves observed at large radii, the small
$\mu_0$ cases of the Misner data actually represent just one single
black hole with non-spherical perturbations, although the spacetime in
the near field region can be very different.  For larger values of
$\mu_0$, we have seen that the initial data sets represent not just
two throats in time symmetric motions.  The waves coming from the past
null infinity and the past horizon play important roles, namely, they
cancel, to a large extent, the radiation emitted before the time
symmetric point of the trajectories.  We have also seen that the
internal dynamics of the black holes do not have much effect on
energies or waveforms observed at large radii.

We are currently extending the semi-analytic approaches to other types
of black-hole events. We feel that such alternative approaches are an
important complement to the direct numerical integration of the
Einstein equations. Although these techniques cannot replace numerical
relativity, which will be the only means of computing detailed
waveforms in more complicated spacetimes such as the full 3D inspiral
of two black holes, they can augment it in a number of ways.  First,
they can be used to test the accuracy of the numerical results, as was
illustrated in this paper. Second, they can predict results well in
advance of the full-blown numerical treatment. Third, they are
inexpensive computationally, and can therefore be used to search the
parameter space (initial separation, angular momenta, mass ratio,
etc.) for potentially interesting phenomena, to be further
investigated using numerical relativity. Fourth, and perhaps most
important, alternative approaches help provide a physical
understanding of the numerical results.

\acknowledgments
We are grateful to Larry Smarr originally suggesting that we undertake
this work.  We are very thankful to Eric Poisson for useful
discussions, and in particular, providing help in the perturbation
calculation in Sec. III.  We thank Steve Brandt for help with data
analysis.  We acknowledge support of National Science Foundation
grants PHY-92-07225, PHY93-96246, PHY94-07882, PHY91-16682,
PHY94-04788, ASC/PHY93-18152 (arpa supplemented). This work was also
supported by research funds of NCSA, the University of Utah, the
Pennsylvania State University and its Office for Minority Faculty
Development.  The calculations were performed at the Pittsburgh
Supercomputing Center and at NCSA.


\begin{references}
\bibitem{beware}
Beware that $M$ denotes the total mass in this paper, while in
Ref.~\cite{Anninos94b}, it is the mass of one hole.
\end{references}
\end{document}